\documentclass[11pt]{article}
\usepackage{mathrsfs}
\usepackage{amsmath}
\usepackage{amsfonts}
\usepackage{amssymb, amsmath, cite}
\usepackage{color}
\usepackage{graphicx}

\usepackage{float}
\usepackage{pgfplots}
\pgfplotsset{compat=newest}

\usepackage{tikz-3dplot}

\newcommand\be{\begin{equation}}
\newcommand\ee{\end{equation}}
\newcommand\ber{\begin{eqnarray}}
\newcommand\eer{\end{eqnarray}}
\newcommand\berr{\begin{eqnarray*}}
\newcommand\eerr{\end{eqnarray*}}
\newcommand\bea{\begin{eqnarray}}
\newcommand\eea{\end{eqnarray}}

\newcommand{\bfR}{{\Bbb R}}
\newcommand{\bfC}{{\Bbb C}}

\newcommand{\x}{{\bf x}}

\newcommand{\lm}{\lambda}

\newcommand{\dd}{\mbox{d}}
\newcommand{\ii}{\mbox{i}}\newcommand{\e}{\mbox{e}}
\newcommand{\pa}{\partial}\newcommand{\om}{\omega}

\newcommand{\nn}{\nonumber}

\newcommand\lb{\label}
\newcommand\eq{\eqref}

\newcommand\sig{\sigma}
\newcommand\HH{\mbox{H}}
\newcommand\sS{\Sigma}
\newcommand\g{{\mbox{g}}}
\newcommand\ou{\overline{u}}

\title{Some Applications of Surface \\Curvatures in Theoretical Physics}

\author{Yisong Yang\\Courant Institute of Mathematical Sciences\\New York University, New York, NY 10012, USA}

\begin{document}
\maketitle

\small{\em On the occasion of the centennial celebration for the launch of mathematics program at Henan University}

\begin{abstract}
In this survey article, we present two applications of surface curvatures in theoretical physics. The first application arises
from biophysics in the study of the shape of cell vesicles involving the minimization of a mean curvature type energy called the
Helfrich bending energy. In this formalism, the equilibrium shape of a cell vesicle may present itself
in a rich variety of geometric and topological characteristics. We first show that there is an obstruction,
arising from the spontaneous curvature, to the existence of a minimizer of the Helfrich energy over the set of embedded ring tori. We then propose a scale-invariant anisotropic bending energy, which extends the
Canham energy, and show that it possesses a unique toroidal energy minimizer, up to rescaling, in all parameter regime.  Furthermore, we establish some
genus-dependent topological lower and upper bounds, which are known to be lacking with the Helfrich energy, for the proposed energy.  We also
present the shape equation in our context, which
extends the Helfrich shape equation. The second application arises from astrophysics in the search for a mechanism for matter accretion in
the early universe in the context of cosmic strings. In this formalism, gravitation may simply be stored over a two-surface so that the Einstein tensor is given in terms of the Gauss curvature of the surface which relates itself directly to the Hamiltonian energy density of the
matter sector. This setting provides a lucid exhibition of the interplay of the underlying geometry, matter energy, and topological characterization of the system. In both areas of applications, we encounter highly challenging nonlinear partial differential equation problems. We demonstrate that
studies on these equations help us to gain understanding of the theoretical physics problems considered.

\medskip


\medskip

\end{abstract}

\section{Introduction}
\setcounter{equation}{0}

Geometric analysis problems arising from physics have always been an inspiring source for the development of many branches of mathematics. 
In this regard, a classical example is Euler's elastica \cite{Tru1,Tru2} whose study has fueled and influenced  subjects such as nonlinear
differential equations, calculus of variations, special functions, and differential geometry. In this context, the central theme of the study is
to understand an equilibrium curve that minimizes Euler's bending energy expressed as an integral of the squared curvature of the curve, evaluated 
against the arclength of the
curve. In this article, we present two problems arising from theoretical physics involving the mean curvature and Gauss curvature of a surface.
The first problem concerns the determination of the equilibrium shape of a cell vesicle by the minimization of a bending energy containing
a weighted sum of the squares of the principal curvatures of the vesicle surface. In its various reduced forms, this energy becomes the 
Willmore energy \cite{MN2,W1}, the Helfrich energy \cite{H}, or the Canham energy \cite{C}. In its general setting, the energy, directly extending
Euler's elastica energy, may be expressed
in terms of a combination of the mean and Gauss curvatures. Our main interests are the existence and nonexistence of the bending energy among
a few geometric and topological conformation classes and topological energy bounds. These are discussed in Sections 2--6.
The second problem arises from the study of cosmic strings and involves only the Gauss curvature of a surface. In this setting, the Einstein tensor
reduces itself into a form that has only two nonzero components, both given in terms of the Gauss curvature of a surface that hosts gravitation.
The Gauss curvature then relates itself to the energy density of the system of the strings through the Einstein equations in such a way that
the problem resembles Nirenberg's problem \cite{Aub} which asks whether a prescribed function can be the Gauss curvature of a conformally deformed surface. In our setting, the prescribed function is the Hamiltonian energy density describing a distribution of strings. Our discussion
will span over a range of models including the Dirac distribution source model of Letelier \cite{Let}, the harmonic map model of Comtet--Gibbons
\cite{CG}, and Lohe's generalized Abelian Higgs model \cite{Lo1} that contains the classical Abelian Higgs model and gauged harmonic map model
as limiting cases. The common features of these constructions are that the Gauss curvature and energy density appear as local lumps as seeds for
matter accretion and these local properties exhibit themselves at infinity forming conical singularity measured by a deficit angle, which also
realizes itself as the total Gauss curvature. These problems are all described by nonlinear partial differential equations similar to that for
Nirenberg's problem. These results and equations are discussed in Sections 7--10.

\section{Motivation in studying bending energies for cell vesicles}
\setcounter{equation}{0}

Cell membranes are essential for life since they separate living cells and their environment and enable metabolism and other life functions to take place.
At the molecular level, a cell membrane is made of phospholipid molecules and an assortment of other molecules such as cholesterol, proteins, and carbohydrates,
to form a fluid mosaic
membrane in the form of a closed surface in the Euclidean space realizing a cell vesicle \cite{SN}. Theoretically,
it is of  importance and interest to understand some universal properties, with regard to geometry, topology, and other mathematical and physical characteristics, of the conformation of a cell vesicle \cite{G,L,St}.
The idea of using curvature bending energies to model the shape of a cell was initiated by  Canham \cite{C} and Helfrich \cite{H}. See \cite{S,Se2,SL} for reviews.
The common feature in various curvature energies is a term that is proportional to the total integral of the square of the mean curvature of the vesicle surface, known as the Willmore energy \cite{MN2,W1}.
On the other hand, the Canham energy density \cite{C} is proportional to the squared sum of the two principal curvatures of the cellular surface.
Thus, in view of the Gauss--Bonnet theorem, the Canham energy differs from the Willmore energy, modulo a constant, by an integral of the Gauss curvature which is a topological invariant. So
the Canham energy is legitimately regarded as to be contained in the Helfrich energy \cite{H} because the energy density of the latter is proportional to
the squared difference of the mean
curvature, and the spontaneous curvature, a quantity taking account of the geometric asymmetry, of the vesicle.
Here, we first show that the spontaneous curvature obstructs the existence  of a solution for the minimization of
the Helfrich energy over the set of embedded ring tori,
except in the Willmore energy situation when the spontaneous curvature vanishes.
Next, we notice that, since the Canham energy term and the integral of the Gauss curvature, in the Helfrich energy, cancel out genus-dependent quantities
and thus conceal the dependence of the total energy on topology, which is also evidenced as in the study of Simon \cite{Si} on the Willmore energy,
the Helfrich energy does not allow an effective capture of the topological information of the vesicle conformation problem.
However, it has long been recognized \cite{DE,F,JL,L,ML,Se,SL} that cells of lipid
bilayers may present themselves in a rich variety of geometric and topological shapes to realize a broad spectrum of life functions.
In particular, vesicles of toroidal as well as high-genus topology are observed  \cite{F,MB,Se,SL}.
Based on these observations, we then identify an appropriate curvature energy which is
consistent with all the well-established curvature energies \cite{L,S,SL,ZK}, and at the same time enables one to effectively extract,
 through direct energy minimization, without obstruction, useful information regarding geometry and topology of a
cell vesicle, as presented in detail in \cite{YPRE}. 

\section{Helfrich curvature energy and obstruction to minimization}
\setcounter{equation}{0}

Let $k_1,k_2$ be the principal curvatures of a closed 2-surface $\Sigma$ immersed in the Euclidean 3-space with area element $\dd \sig$.
The well-known Helfrich curvature
bending energies modeling the shape of a cell vesicle is \cite{H}:
\be\label{1}
U_{\HH}(\Sigma)=\int_\Sigma \frac12\kappa \left(k_1+k_2-c_0\right)^2\,\dd \sigma,
\ee
where the constants $\kappa>0$ is the bending modulus and
$c_0$  the spontaneous curvature dictating the asymmetry or bending tendency of the surface. When $c_0=0$, (\ref{1}) is the well-known Willmore
energy in differential geometry \cite{W1}. 

{\bf Obstruction to existence of energy minimizer}

To see how $c_0$ arises as an obstruction to the existence of a minimizer of (\ref{1}), we consider the set
of embedded ring tori $\{T^2_{a,b}\}_{a>b>0}$ where $T^2_{a,b}$ is defined by the parametrization
\be
{\bf x}(u,v)=\left((a+b\cos u)\cos v,(a+b\cos u)\sin v, b\sin u\right),\quad 0\leq u, v\leq 2\pi,
\ee
as plotted in Figure \ref{F1}, where ${\bf x}(u,v)=(x(u,v),y(u,v),z(u,v))$.
\begin{figure}[H]
\begin{center}
\begin{tikzpicture}[scale=2]
  \begin{axis}[
      axis equal image,
      axis lines=middle,
      xmax=18,zmax=10,
      ticks=none,
      clip bounding box=upper bound,
      colormap/blackwhite
    ]

    \addplot3[domain=0:360,y domain=0:320, samples=50,surf,z buffer=sort]
    ({(12 + 3 * cos(x)) * cos(y)} ,
    {(12 + 3 * cos(x)) * sin(y)},
    {3 * sin(x)});
    \draw [thick,blue] (axis cs: 0,0,0) -- (axis cs: 12,0,0) node [midway,above=-2] {$a$};
    \draw [thick,red] (axis cs: 12,-0.2,0) -- (axis cs: 12,3.7,0) node [midway,below right=-3] {$b$};

    \draw [-latex] (axis cs: 0,0,0) -- node [pos=0.9, xshift=0.5em]{$z$}(axis cs: 0,0,10);
    \draw [-latex] (axis cs: 0,-15,0) --
    node [pos=0.9, xshift=-1em, yshift=0.5em]{$x$}(axis cs: 0,-20,0);
    \draw (axis cs: 0,0,0) -- (axis cs: 0,9,0);
    \draw (axis cs: 0,0,0) -- (axis cs: -9,0,0); 
\draw [-latex] (axis cs: 10,0,0) -- node [pos=0.9, xshift=0.75em]{$y$}(axis cs: 18,0,0);
  \end{axis}
\end{tikzpicture}
\end{center}
\caption{An illustrative plot of an embedded totus in $\bfR^3$, denoted by $T^2_{a,b}$, where $a$ is the distance from the center of the tube of the
torus to the center of the torus and $b$ the radius of the tube. }\lb{F1}
\end{figure}
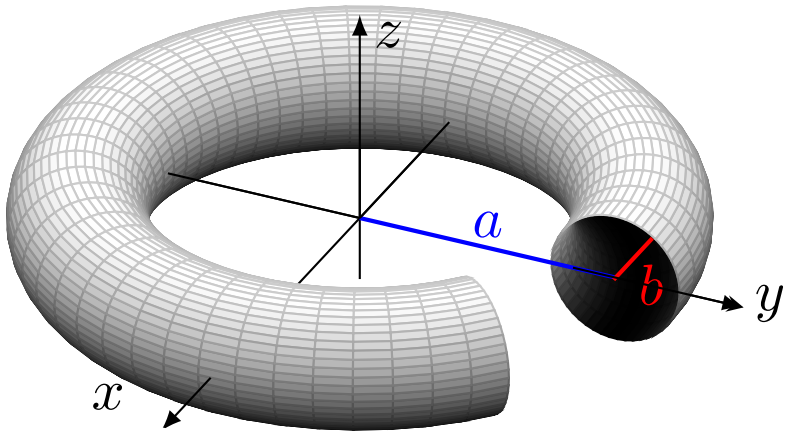
Thus the principal curvatures and area element of $T^2_{a,b}$ are
\be
k_1=-\frac1b,\quad k_2=-\frac{\cos u}{a+b\cos u},\quad \dd\sig=(a+b\cos u)b\,\dd u\dd v.
\ee
Inserting these into (\ref{1}) and using the ratio of the generating radii, $\tau=b/a$, as a new variable, we get
\be\label{2}
 U_{\HH}(T^2_{a,b})={2\pi^2\kappa}\left(\frac1{\tau\sqrt{1-\tau^2}}+2c_0 a+c_0^2 a^2\tau \right),
\ee
where $a>0,\tau\in(0,1)$.
When $c_0>0$, the infimum of (\ref{2}) is attained at $\tau=1/{\sqrt{2}}, a=0$, which is $4\pi^2\kappa$. Thus we see that this infimum is {\em not}
attainable, that is, the Helfrich energy (\ref{1}) cannot be minimized, among the ring tori. When $c_0=0$, the infimum of (\ref{2}) is attainable
for any $a>0$ and $\tau=1/\sqrt{2}$, which is the classical Willmore situation \cite{W1}. When $c_0<0$, we see that for fixed $\tau\in(0,1)$ the right-hand side of (\ref{2}) can be minimized at $a=1/{|c_0|\tau}$. For such a choice of $a$, we obtain from (\ref{2}) the result
\be
 U_{\HH}(T^2_{a,b})=\frac{2\pi^2\kappa(1/\sqrt{1-\tau^2}-1)}{\tau},
\ee
which is monotone increasing and tends to zero as $\tau\to0$. Thus,
the infimum of (\ref{2}) is zero which is again not attainable among the ring tori
considered. In other words, (\ref{1}) allows a minimizer over the set of embedded ring tori if and only if $c_0=0$, which is the Willmore limit,  indicating that the
 spontaneous curvature presents an obstruction to the minimization of the Helfrich energy (\ref{1}) over ring tori.

{\bf Full Helfrich bending energy and relaxation of obstruction}

Furthermore, we consider the Helfrich bending energy in its general form \cite{H,ZH1} containing contributions from the volume and surface area of the cell vesicle:
\be\label{3}
{\cal F}_{\HH}(\Sigma)=U_{\HH}(\Sigma)+p\int_{\cal V}\dd v+\lm\int_{\Sigma}\dd\sig,
\ee
where $\cal V$ is the cellular region enclosed by the vesicle $\Sigma$, with $\dd v$ the volume element, and $p$ and $\lm$, respectively, are the osmotic pressure difference between the inside and outside the cell membrane and the surface tension. Physically, the lipid bilayer structure of the cell membrane results in a one-way
traffic flow of salt, allowing salt to enter the cell but not leak away, which leads to a jump of salt concentration and hence a positive pressure difference, $p>0$.
On the other hand, surface tension of the plasma membrane of the cell dictates an elastic preference for the vesicle to assume as small as possible a surface area,
thus leading to $\lm>0$ as well. In our study here, we will observe these non-degenerate restrictions. Recall that, for $\Sigma=T^2_{a,b}$, we have the volume
$V=\int_{\cal V}\dd v=2\pi^2 ab^2$ and surface area $A=\int_{\Sigma}\dd\sigma=4\pi^2 ab$ for
the toroidal vesicle concerned. Hence, from (\ref{3}), we see that (\ref{3}) becomes
${\cal F}_{\HH}(T^2_{a,b})=2\pi^2 h(a,\tau)$ where
\be\label{4}
h(a,\tau)=\frac{\kappa}{\tau\sqrt{1-\tau^2}}+2\kappa c_0 a+\left(\kappa c_0^2+2\lm\right) a^2\tau+p a^3\tau^2.
\ee
When $c_0\geq0$, the infimum of this function is obtained by setting $a=0,\tau=1/\sqrt{2}$. Thus (\ref{3})  has no minimum
over the set of ring tori. When $c_0<0$, we set ${\pa h}/{\pa a}=0$ to get the solution
\be\label{5}
a_0=\frac{\sqrt{(\kappa c_0^2+2\lm)^2-6\kappa c_0 p}-(\kappa c_0^2+2\lm)}{3p\tau},
\ee
and $b_0=a_0\tau$ for the generating radii in terms of the coupling parameters and the ratio parameter $\tau\in(0,1)$. In view of these, we
deduce that $h(a,\tau)$ has a unique minimizer with $a=a_0$ given earlier and $\tau=\tau_0$ which is the unique root of the equation $h_1'(\tau)=0$, where
$h_1(\tau)=h(a_0,\tau)=h(b_0/\tau,\tau)$, which may be reduced to
\be\label{6}
\frac{1-2\tau^2}{(1-\tau^2)^{\frac32}}=\frac{b_0^2}\kappa\left([\kappa c_0^2+2\lm]+2pb_0\right),
\ee
for $\tau\in(0,1)$. A direct consequence of interest from (\ref{6}) is the following {universal}, parameter-independent,  bounds for the ratio $\tau_0$,
\be\label{7}
0<\tau_0<\frac1{\sqrt{2}}.
\ee
In addition, of independent interest is that we may use (\ref{7}) to obtain a sharpened estimate for $\tau_0$. In fact, we notice that the quantity on the right-hand
side of (\ref{6}) lies in the interval $(0,1)$, which allows us to express it as $1-\beta_0$. Thus, in view of (\ref{7}), we may infer by (\ref{6}) that $\tau_0$
satisfies the strengthened parameter-dependent bounds
\be\label{8}
\sqrt{1-\frac1{1+\beta_0}}<\tau_0<\sqrt{1-\frac1{1+\sqrt{\beta_0}}}.
\ee
It is worth noting that the classical Willmore ratio, $1/{\sqrt{2}}$, would appear in the limit $\beta_0\to1$ in (\ref{8}), but would
actually never happen
for any concrete choice of the coupling parameters, indicating the phenomenological richness of the geometric content included in (\ref{3}).
Besides, it is easily seen that (\ref{3}) has no minimizer over the set of spheres when $c_0\leq0$, but has a unique minimizer whose radius
is given by 
\be
R_0=\frac{\sqrt{(2\lm+\kappa c_0^2)^2+8\kappa p c_0}-(2\lm+\kappa c_0^2)}{2p},
\ee
 when $c_0<0$.
Thus, we see that (\ref{3}) relaxes the obstruction to
the existence of a minimizer presented by the spontaneous curvature $c_0$, and that, in (\ref{3}), topology plays a role in selecting the sign of $c_0$,
through minimization.  However, in both the spherical and toroidal cases, scale invariance is broken, which may be regarded as another type of obstruction
to existence.
We may compare our toroidal results here on the minimization of the vesicle energy (\ref{3}) over the set of ring tori with those obtained in \cite{Mutz,Zhong} where
  it is found that
$c_0$ in the bending energy must stay negative for the existence of a stable toroidal vesicle, which is consistent with our results, and that the
ratio of the generating radii of the toroidal vesicle takes the Willmore value $1/\sqrt{2}$, which is inconsistent with our findings, (\ref{7}) and (\ref{8}).

{\bf Lack of topological bounds}

After an illustration on the obstruction to the existence of an energy minimizer, and its relaxation, arising from the spontaneous curvature in (\ref{1}) and (\ref{3}), it is of interest to briefly discuss
the lack-of-topological-bound problem associated.  In fact, when $c_0=0$, (\ref{1}) is the Willmore energy which has the classical lower bound $8\pi\kappa$, independent of the genus of $\sS$. Besides, by the study of
Simon \cite{Si}, we know that the infimum of this Willmore energy actually lies in the interval $[8\pi\kappa, 16\pi\kappa)$,
which is also genus independent. When $c_0\neq0$, the difference between
the Willmore energy and (\ref{1}) or (\ref{3}) is a sum of the quantities involving the average of the mean curvature, surface area, and volume, of the vesicle,
which is non-topological. 

\section{Bending energy based on principal curvatures}
\setcounter{equation}{0}

To motivate the introduction of the energy functional of our study, we first recall the Canham vesicle energy \cite{C}:
\be
U_{\mbox{\small C}}(\Sigma)=\int_{\Sigma}\frac12\kappa (k_1^2+k^2_2)\,\dd\sigma,
\ee
proposed to describe the observed biconcave shape of a red blood cell, which differs from the Willmore energy only by a topological invariant.
 We then recall a more general membrane-bending energy
\be
U_{\mbox{\small M}}(\sS)=\int_{\sS}\left(\frac12\kappa_+\left(k_1+k_2\right)^2+\frac12\kappa_-\left(k_1-k_2\right)^2\right)\dd\sigma,
\ee
which describes the shape a biological fluid membrane  in \cite{GK,M,D}
subject to a thermal environment, where $\kappa_+$ and $\kappa_-$ are two elastic moduli incorporating the asymmetric bending tendency as a consequence of
thermal fluctuations.

{\bf Bending energy involving principal curvatures}

Combining and balancing the isotropic feature of $U_{\mbox{\small C}}(\sS)$ and anistropic ingredient in $U_{\mbox{\small M}}(\sS)$, we are led to considering the bending energy
\be\label{11}
U(\sS)=\int_{\sS}\frac12\left(\kappa_1 k_1^2+\kappa_2 k_2^2\right)\,\dd\sigma,
\ee
where $\kappa_1,\kappa_2>0$ are two bending rigidities or moduli, included to embrace a wider range of possible anisotropic phenomenology for lipid bilayer surfaces.
 Bear in mind that the principal curvatures $k_1,k_2$ may be represented in terms of
the mean and Gauss curvatures $H=(k_1+k_2)/2$ and $K=k_1 k_2$,  of the surface, by
$
\left\{k_1,k_2\right\}=\left\{H+\sqrt{H^2-K},H-\sqrt{H^2-K}\right\}.
$
Thus, it is instructive to see that  (\ref{11}) may be rewritten as
\be\label{12}
U(\sS)=\int_\sS\left(\om H^2-\frac{\om}2 K\pm\delta H\sqrt{H^2-K}\right)\,\dd\sig,
\ee
where $\om=\kappa_1+\kappa_2$ and $\delta=|\kappa_1-\kappa_2|$ are the sum and absolute difference of the elastic moduli, respectively,
 the first term is the Willmore energy density,  the second  the Gauss--Bonnet topological invariant density,
and the third a new quantity taking account of the anisotropy of the bending energy.  Here the sign convention in (\ref{12}) follows the
rule that the plus sign is chosen when the greater
bending rigidity is associated to the greater principal
curvature and the negative sign is chosen when the greater bending rigidity is associated to the smaller principal curvature. It is clear that the
role of the third term when $\delta\neq0$
works to signal out the presence of the non-umbilicity of the surface and break the democracy between the principal curvatures. Thus,
in  the context of the model (\ref{11}), anisotropy replaces the role of the spontaneous curvature  in the Helfrich energy (\ref{1}), so that
a broader range of phenomenology may be achieved.

\section{Existence of energy minimizer and topological bounds} 
\setcounter{equation}{0}

We now present the anticipated properties of the energy (\ref{11}) or (\ref{12}), lacking with (\ref{1}) and (\ref{3}), namely, the existence of a unique
energy minimizer over the set of embedded ring tori, and $\g$-dependent lower and upper bounds of the energy, in its full parameter regime.
See Figure \ref{F2} for the illustration of a genus one surface and explanation how a higher genus surface may be constructed systematically.

{\bf Existence and uniqueness of energy minimizer}

First, using the parametrization of the ring torus $T^2_{a,b}$ considered earlier and setting $\gamma=\kappa_1/\kappa_2,\kappa_2=\kappa$, we see
that (\ref{11}) may be evaluated to give us the scale-invariant quantity
$
U(T_{a,b}^2)=2\pi^2\kappa f(\tau)$, where
$f(\tau)=\gamma/\tau+\tau/{\sqrt{1-\tau^2}(1+\sqrt{1-\tau^2})}$.
Since $f(\tau)\to\infty$ as $\tau\to0$ and $\tau\to1$, we see that $f(\tau)$ attains its global minimum in $0<\tau<1$, for any $\gamma>0$,
which is a root of $f'(\tau)=0$ in $(0,1)$, which happens to be unique, and may be denoted as $\tau_{\min}$, which satisfies
the simplified equation
\be\label{14}
\gamma=1+\frac{2\tau^2-1}{(1-\tau^2)^{\frac32}}.
\ee
Given $\gamma$, we can write $\tau_{\min}$ in a closed-form expression
 which in general is rather complicated. Nevertheless the monotone dependence of $\tau_{\min}$ on $\gamma$ is clear by
the implicit function theorem such that $\tau_{\min}\to 1$ when $\gamma\to\infty$ and $\tau_{\min}\to0$ when $\gamma\to 0$. As some concrete examples, we take $\gamma$ to be
$\gamma=N$ ($N=1,2,\dots,100$), and get the following results which are sufficiently simple to be listed
for the pair $\left(\tau_{\min},\gamma\right)$:
$\left(1/{\sqrt{2}},1\right),
\left({\sqrt{3}}/2,5\right),\left({2\sqrt{2}}/3,22\right),\left({\sqrt{15}}/4,57\right),$
among which $\left(1/{\sqrt{2}},1\right)$ is the classical result in the Willmore problem \cite{MN2,W1}.  On the other hand, (\ref{14})
allows us to find $\gamma$ easily for prescribed $\tau_{\min}$.
Thus, given arbitrary $\kappa_1,\kappa_2$, we can insert (\ref{14}) with $\tau=\tau_{\min}$ into $U(T^2_{a,b})=2\pi^2\kappa f(\tau)$ to determine the minimum value of the bending energy (\ref{11}) over the set of  ring tori  to be
\be\label{15}
U_{\min}(T^2)=2\pi^2\kappa\frac{\tau_{\min}}{(1-\tau_{\min}^2)^{\frac32}}.
\ee
For example, $U_{\min}(T^2)=4\pi^2 \kappa$ when $\gamma=1$ (so that $\tau_{\min}=1/\sqrt{2}$), which is classical, and
$
U_{\min}(T^2)=8\pi^2\kappa/{3\sqrt{3}}$ when $ \gamma=1-4/{3\sqrt{3}}$ (so that $\tau_{\min}=1/2$), say.

{\bf Topological bounds}

We now obtain some topological lower and upper bounds for the bending energy (\ref{11}) or (\ref{12}).
The basic quantity that concerns us is
$
U_{\g}=\inf\left\{ U(\sS)\,|\,\mbox{$\sS$ is of genus $\g$}\right\}.
$
To proceed,
we may apply
the Chern--Lashof inequality \cite{CL}
\be
\int_\sS|K|\,\dd\sig\geq 4\pi(1+\g)
\ee
to derive the result
\be\label{16}
U(\sS)\geq\sqrt{\kappa_1\kappa_2}\int_{\sS}|K|\,\dd\sigma
\geq 4\pi(1+\g)\sqrt{\kappa_1\kappa_2},
\ee
which is $\g$-dependent as desired.
From (\ref{16}) we have the lower bound
$
U_\g\geq 4\pi(1+\g)\sqrt{\kappa_1\kappa_2}.
$
We now turn our attention to obtaining some $\g$-dependent upper bounds for $U_\g$ ($\g=0,1,2,\dots$).
First consider $\g=0$. In this situation, we use the 2-sphere of radius $R>0$, say $S^2_R$, as a trial surface. Then $k_1=k_2=1/R$ so that $U(S^2_R)=2\pi (\kappa_1+
\kappa_2)$. Thus we have
$
\sqrt{\kappa_1\kappa_2}\leq{U_0}/{4\pi}\leq{(\kappa_1+\kappa_2)}/2.
$
That is, the quantity ${U_0}/{4\pi}$ lies between the geometric mean and the arithmetic mean of the bending moduli. In particular, in the isotropic limit,
$\kappa_1=\kappa_2=\kappa$ (say), $U_0=4\pi\kappa$, which is realized by all round spheres. This last statement is a classical result due to Willmore \cite{W1}.
In the anisotropic situation where $\kappa_1\neq\kappa_2$, it is inevitable to anticipate that $U_0$  be realized by non-round spheres such as ellipsoidal or biconcave
surfaces due to the structure of the energy (\ref{11}).
Next, consider $\g=1$. In view of (\ref{15}), we have
$
U_1\leq U_{\min}(T^2)= 2\pi^2\kappa{\tau_{\min}}/{(1-\tau^2_{\min})^{\frac32}}$ (equality is true when $\gamma=1$ \cite{MN2}).
For $\g=2$, let $T^2_{\min}$ be a ring torus realizing the minimum energy $U_{\min}(T^2)$ given in (\ref{15}) and suitably glue two copies of $T^2_{\min}$ together to get a smooth $\g=2$ surface, say $\tilde S$, which may be made to satisfy
$
U({\tilde S})\leq 2 U_{\min}(T^2).
$
 Hence $U_2\leq 2U_{\min}(T^2)$. Extending this argument, we establish the
general bound $U_{\g}\leq \g\, U_{\min}(T^2)$ for any $\g\geq2$.
Summarizing,
 if we denote the unique solution $\tau_{\min}$ of the equation (\ref{14}) by $\tau(\gamma)$, then we
arrive at the $\g$-dependent bounds
\be\label{18}
(1+\g)\sqrt{\gamma}\,\kappa\leq \frac{U_\g}{4\pi}\leq \frac{{\g \pi\kappa}\tau(\gamma)}{2(1-\tau^2(\gamma))^{\frac32}},\, \,\,\g\geq1.
\ee
In particular, in the isotropic situation when $\gamma=1$ so that $\tau(1)=1/{\sqrt{2}}$, the bounds stated in (\ref{18}) assume the following elegant simple form:
\be\label{19}
(1+\g)\,\kappa\leq \frac{U_\g}{4\pi}\leq \g\pi  \kappa,\, \,\,\g\geq1,
\ee
among which the case when $\g=1$ is classical and the right-hand-side inequality is actually equality \cite{MN2}.

\begin{figure}[H]
\begin{center}
    \tdplotsetmaincoords{55}{110}
    \begin{tikzpicture}[tdplot_main_coords]
    \path[tdplot_screen_coords,use as bounding box] (-3.2,-3.2) rectangle (6,6);
    \pgfmathsetmacro{\R}{3}
    \pgfmathsetmacro{\myang}{20}
    \pgfmathsetmacro{\angtop}{-40}
    \pgfmathsetmacro{\angright}{95}
    \coordinate (O) at (0,0,0);
    \shadedraw [ball color=brown!50,tdplot_screen_coords] (0,0,0) circle(\R);
    \path 
    ({3*sin(\myang)*cos(\angright)},{3*cos(\myang)},{3*sin(\myang)*sin(\angright)}) coordinate (P1)
    ({3*sin(\myang)*cos(\angtop)},{3*sin(\myang)*sin(\angtop)},{3*cos(\myang)}) coordinate (P2);
    \draw[right color=white,left color=orange!100,shading angle=300] plot[variable=\x,domain=\angtop:\angtop+180,samples=91]
    ({3*sin(\myang)*cos(\x)},{3*sin(\myang)*sin(\x)},{3*cos(\myang)})
    to[out=90,in=0,looseness=2] (P1)
    plot[variable=\x,domain=\angright:\angright-180,samples=91]
    ({3*sin(\myang)*cos(\x)},{3*cos(\myang)},{3*sin(\myang)*sin(\x)})
    to[out=0,in=90,looseness=4] (P2);
    \end{tikzpicture}
\caption{A genus one surface viewed as a sphere with a handle attached to it which is topologically identical to the embedded torus shown in
Figure \ref{F1}. A surface with genus $\g=n$ may be produced this way by attaching $n$ handles to a sphere. }\lb{F2}
\end{center}
\end{figure}
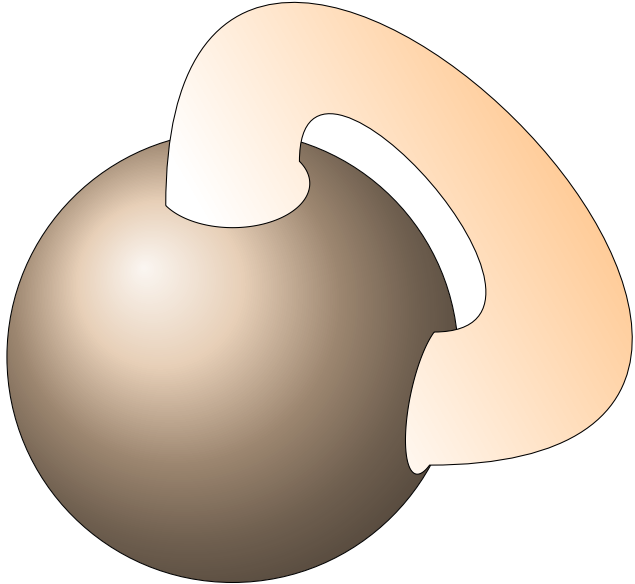

\section{Shape equation of bending energy}
\setcounter{equation}{0}

We can now present the shape equation
of the anisotropic vesicle bending energy (\ref{11}) or (\ref{12}). Here, for greater generality and applicability,
we consider instead the following shape energy,
extending (\ref{3}), as proposed in \cite{H}, along the study of Ou-yang and Helfrich
\cite{ZH1}:
\be\label{20}
{\cal F}(\sS)=U(\sS)+\int_{\sS}\left(\xi_1 k_1+\xi_2 k_2+\lm\right)\dd\sig+p\int_{\cal V}\dd v,\quad
\ee
where $\xi_1,\xi_2,p,\Lambda$ are suitable parameters.
 By direct variation of (\ref{20}), we arrive at its shape equation:
\bea\label{21}
&& p-2\Lambda H \mp\frac12\,\Delta\left(\frac {\delta K+\zeta H}{\sqrt{H^2-K}}\right)
\pm\frac12\, M\left(\frac {\delta H+\zeta}{\sqrt{H^2-K}}\right)\nn\\
&&+\om\Delta H-2H^2\left(\om H+2\xi \right)
+2(\om H+\xi) (2H^2-K)\nn\\
&&\pm\frac{\delta}{\sqrt{H^2-K}}\left(2H^4-3H^2 K+K^2\right)=0,
\eea
where ${\xi}=(\xi_1+\xi_2)/2$,  $\zeta=\xi_1-\xi_2$, $\Delta$ is the Laplace--Beltrami operator, and the operator $M$ is defined by 
\bea
\sqrt{EG}M\eta&=&
\left(k_1\sqrt{\frac GE}\eta\right)_{uu}+\left(k_2\sqrt{\frac EG}\eta\right)_{vv}\nn\\
&&+\frac12\left(\left[k_1\frac{E_u}E\sqrt{\frac GE}\eta\right]_u
-\left[k_1\frac{E_v}{\sqrt{EG}}\eta\right]_v
+\left[k_2\frac{G_v}G\sqrt{\frac EG}\eta\right]_v-\left[k_2\frac{G_u}{\sqrt{EG}}\eta\right]_u\right),
\eea
in curvature coordinates (or lines of curvature), where $E,F,G$ are the coefficients of the first fundamental form of $\sS$, with $F=0$.
In the Helfrich isotropic limit \cite{H} where the leading term of the energy is given as in (\ref{1}), we have $\kappa_1=\kappa_2=\kappa, \delta=0,\xi_1=\xi_2=-\kappa c_0$ ($c_0$ being the spontaneous curvature), $\zeta=0$, $\om=2\kappa,\xi=-\kappa c_0$, and
$
\Lambda=\lm+\kappa c_0^2/2,
$
where $\lm$ is the surface tension of the cell membrane as in (\ref{3}). Thus the equation (\ref{21}) becomes the classical shape equation \cite{ZH1}
\be\label{22}
p-2\lm H+\kappa(2H-c_0)(2H^2-2K+c_0 H)+2\kappa \Delta H=0.
\ee

It has also been shown that anisotropy of the bending energy (\ref{11}) naturally allows a broad range of phenomenology for the shaps of a vesicle. For example, when $\g=0$,
ellipsoidal and biconcave surfaces indeed occur as
energetically
favored geometries over a round sphere \cite{YPRE}. Here we omit the detailed discussion.

\medskip

In summary, we have seen that the spontaneous curvature in the Helfrich bending energy obstructs its minimization and that, like
the Willmore energy, it lacks
topology-dependent energy bounds.
We have shown that these difficulties can be overcome by using a scale-invariant anisotropic curvature energy extending that of Canham as the bending energy for the shape of a cell vesicle
so that its minimization
over the set of embedded ring tori  always has a unique solution, up to rescaling, for arbitrary choice of the parameters, and displays a clear transition of various
geometric shapes of a vesicle, and that the energy stays between some natural topology-dependent lower and upper bounds expressed linearly in terms of the genus of the vesicle. The study here offers rich opportunities for the phenomenological study of the geometric and topological characteristics of a cell vesicle \cite{YPRE}.

\section{Einstein tensor and Gauss curvature}
\setcounter{equation}{0}

The essence of quantum field theory is the use of gauge fields that arise to restore local symmetry of the underlying matter field theory
with a characteristic global internal symmetry such that the symmetry group $U(1)$ gives rise to electromagnetic forces, $SU(2)$ to weak forces,  $SU(3)$
to strong forces, and $SU(5)$ to a grand-unified theory. With gauge fields, the conventional partial derivatives operating on matter fields are replaced by gauge-covariant derivatives, under the notion of connection, and the commutators of these derivatives are then measured by various field
strength tensors, or curvatures. On the other hand, Einstein's gravity theory, or general relativity, is based on preserving local symmetry of spacetime 
inherited from
an external global symmetry, or the Lorentz symmetry, of the flat spacetime when gravity is absent. To achieve this goal, it is realized that
the presence of gravity leads to formulating the spacetime with a 4-dimensional Riemannian manifold with a metric element \cite{Dir}, say
\be\lb{x1}
\dd s^2=g_{\mu\nu}\dd x^\mu\dd x^\nu,\quad \mu,\nu=0,1,2,3,
\ee
with the Minkowski signature $(+---)$, in local coordinates and assuming summation convention over repeated indices, such that $\mu=0$ corresponds to temporal and $\mu=i=1,2,3$ to space coordinates. As
a consequence of preserving local symmetry, covariant derivatives, $\nabla_\mu$, defined by the associated Riemannian connection now replace the conventional
partial derivatives, whose noncommutativity is thereby measured by a mixed tensor field, written $R^\alpha_{\mu\nu\beta}$, called the Riemann tensor. 
The Ricci tensor $R_{\mu\nu}$ comes up by contracting the Riemann tensor, $R_{\mu\nu}=R^\alpha_{\mu\nu\alpha}$, which in turn gives rises to
the scalar curvature $R=g^{\mu\nu}R_{\mu\nu}$ (with the notation $(g^{\mu\nu})=(g_{\mu\nu})^{-1}$). Einstein's idea of gravitation is to
find a suitable geometric quantity that is made proportional to the usual energy-momentum tensor, $T_{\mu\nu}$, of the matter (physical) content in
the spacetime. Since $T_{\mu\nu}$ is assumed to be a conserved quantity in view of Noether's theorem, that is, it is divergence-free with respect to
covariant (or rather, contravariant) derivatives, the desired geometric quantity must also be a divergence-free quantity of the same tensor type as
$T_{\mu\nu}$. Based on compatibility, generality, simplicity, and dynamical property considerations, Einstein found the unique choice \cite{Dir}
\be\lb{x2}
G_{\mu\nu}=R_{\mu\nu}-\frac12 g_{\mu\nu}R,
\ee
which is called the Einstein tensor, which enabled him to arrive at the relation $G_{\mu\nu}=-\kappa T_{\mu\nu}$ as envisioned, where $\kappa$ is
a proportionality constant. It can shown that, in order to recover Newton's law of gravity asymptotically in weak-field limit, $\kappa$ should
read $\kappa=8\pi G$, where $G$ is Newton's universal gravitational constant which is tiny. In conclusion, we have just quickly completed the journey of
Einstein to arrive at his celebrated gravitational equation
\be\lb{x3}
G_{\mu\nu}=-8\pi GT_{\mu\nu},
\ee
called the Einstein equations, which is a system of nonlinear second-order partial differential equations in the unknowns $g_{\mu\nu}$
and takes the form of coupled wave equations in weak-field limit, thus predicting the occurrence of gravitational waves. Due to the nonlinearity involved,
these equations are notoriously difficult in their general setting. However, when some of their highly specialized and simplified settings are considered, 
mathematicians and theoretical physicists have harvested fruitfully. For example, the purely time-dependent but spatially-independent case of (\ref{x3})
forms the foundation of modern theory of cosmology and time-independent but spatially-dependent and spherically symmetric case of (\ref{x3}) leads to
the conceptualization of black holes. See \cite{Car,MTW,Wald}. In this part of the article, we consider, yet, another specialized and simplified case
of (\ref{x3}), involving the Gauss curvature of a two-surface, which is also related to an interesting geometric analysis problem called Nirenberg's problem,
which we now review briefly in our context.

{\bf Nirenberg's problem}

For a recent study of Nirenberg's problem, see \cite{Anderson,Aub}. The original problem was formulated over the unit sphere $S^2$. Here, for our purposes, we
consider its generalized setting over a two-dimensional surface $S$ without boundary equipped with a metric form $g=(g_{ij})$, also collectively denoted by $(S,g)$.

Let $g_0$ and $ g$ be two metrics on $S$ and $K_{g_0}$ and $ K_g$ the associated Gauss curvatures, respectively. If $g_0$ and $g$ are related conformally through
the pointwise expression
\be\lb{x4}
g=\e^\eta g_0,
\ee
where $\eta$ is a function over $S$, then $K_{g_0}$ and $K_g$ are related by the equation
\be\lb{x5}
-\Delta_{g_0}\eta+2K_{g_0}=2K_g \e^{\eta},
\ee
where $\Delta_g$ is the usual Laplace--Beltrami operator with respect to $g$ defined by
\be
\Delta_g \eta=\frac1{\sqrt{\det(g)}}\pa_i\left(g^{ij}\sqrt{\det(g)}\pa_j\eta\right),\quad i,j=1,2.
\ee
Nirenberg's problem asks: Given $(S,g_0)$ and a scalar function $F(x)$ over $S$, can one find a conformal deformation of $g_0$, namely $g$ as described in (\ref{x4}), such that the Gauss curvature $K_g$ of $(S,g)$ is exactly the function $F(x)$, or $K_g=F$ over $S$? In view of the relation (\ref{x5}),
this problem amounts to knowing whether the equation
\be\lb{x7}
-\Delta_{g_0}\eta+2K_{g_0}=2F(x) \e^{\eta}
\ee
has a solution, which has not been fully understood yet (see comments in \cite{Anderson,Aub} and references therein). To gain some insight to this problem,
we assume that $S$ is compact. Integrating (\ref{x7}) over $(S,g_0)$ and using the Gauss--Bonnet theorem, we obtain
\be\lb{x8}
\int_S F(x)\e^\eta\dd\Omega_{g_0}=\int_S K_{g_0}\dd\Omega_{g_0}=2\pi \chi(S),
\ee
where $\chi(S)=2-2\g$ is the Euler characteristic of $S$ with $\g$ being the genus of $S$, so that, in the original Nirenberg's problem, $S=S^2, \g=0, K_{g_0}=1$. It may be checked that the existence of a solution to (\ref{x7}) subject to (\ref{x8}) is equivalent to the solvability of the minimization
problem
\be\lb{x9}
\min\left\{\int_S\left(\frac12|\nabla \eta|^2+2K_{g_0}\eta\right)\dd\Omega_{g_0}\,\bigg|\int_S F(x)\e^\eta\dd\Omega_{g_0}=2\pi\chi(S)\right\},
\ee
where $|\nabla\eta|^2=\pa_i\eta \pa^i\eta=g^{ij}\pa_i\eta\pa_j\eta$. Besides, we may also decompose $\eta$ in (\ref{x7}) as
\be\lb{x10}
\eta=c+u,\quad c\in\bfR,\quad \int_S u\,\dd\Omega_{g_0}=0.
\ee
In view of (\ref{x10}) and solving for $c$ from (\ref{x8}), we obtain the following non-local equation,
\be\lb{x11}
-\Delta_{g_0}u+2K_{g_0}=\frac{4\pi\chi(S) F(x) \e^{u}}{\int_S F(x)\e^u \,\dd\Omega_{g_0}},\quad \int_S u\,\dd\Omega_{g_0}=0,
\ee
which is sometimes referred to as a mean-field equation, due to its occurrence in the mean-field theory one-body approximation of a many-body system
by methods of statistical mechanics. Alternatively, one may also consider a heat-flow equation associated with (\ref{x7}), with unknown
$\eta=\eta(t,x)$, given by
\be\lb{x12}
\frac{\pa\eta}{\pa t}-\Delta_{g_0}\eta+2K_{g_0}=2F(x) \e^{\eta},\quad \eta(0,x)=\eta_0(x),
\ee
where $\eta_0$ is an initial function. Hopefully, when $\eta_0$ is suitably chosen, the solution $\eta=\eta(t,x)$ of (\ref{x12}) will approach an
equilibrium state as $t\to\infty$ in a certain sense so that it yields a solution to (\ref{x7}). Since it is clear that the functional
\be\lb{x13}
I(\eta)=\int_S\left(\frac12|\nabla \eta|^2+2K_{g_0}\eta-2F(x)\e^\eta\right)\dd\Omega_{g_0}
\ee
decreases along the flow of (\ref{x12}), the method here could analytically resemble a minimization approach.

{\bf Einstein tensor, cosmic strings, and Gauss curvature}

Soliton-like structures such as domain walls, vortices, monopoles, and instantons have played important roles in fundamental physics for
over a half century. They occur as a consequence of spontaneous symmetry breaking in internal spaces and often are characterized as topological
defects. When gravity is considered, such solitons, exhibiting themselves as energy lumps, clearly cause the spacetime geometry to inherit
such properties as demonstrated by the Einstein equations (\ref{x3}). More precisely, it is natural to anticipate the geometry of the spacetime to
``curl up" at such energy lumps, which would give rise to a possible mechanism for
the appearance of sites as seeds for matter accretion or accumulation in the early universe. Specifically, vortex lines have been used in cosmology
to generate large-scale string structures, called cosmic strings \cite{Gar,Gre,K,V,VS,Witten}. In this context, the fields involved are static and
enjoy an axial-symmetry, that is, they are homogeneous along a fixed coordinate direction in space, namely the $x^3$ coordinate axis in
Cartesian coordinates. Accordingly, the simplest form of the gravitational metric element, (\ref{x1}), may consistently be given by the expression \cite{CG}
\be\lb{x14}
\dd s^2=\dd t^2 -(\dd x^3)^2-\e^\eta((\dd x^1)^2+(\dd x^2)^2),\quad \eta=\eta(x^1,x^2).
\ee
In other words, the spacetime manifold is taken to be the curled Minkowski space $\bfR^{1,3}\approx \bfR^{1,1}\times (\bfR^2,\e^\eta \delta_{ij})$ where nontrivial
geometry or gravitation is assumed to be contained in the conformally flat 2-surface $(\bfR^2,\e^\eta \delta_{ij})$. In view of (\ref{x14}), a
direct computation gives us the components of the Einstein tensor (\ref{x2}) as follows,
\be\lb{x15}
G_{00}=-G_{33}=\frac12 \e^{-\eta}\Delta\eta;\quad G_{\mu\nu}=0,\quad (\mu,\nu)\neq(0,0)\mbox{ or } (3,3).
\ee
However, from (\ref{x5}), we see that the Gauss curvature, denoted now as $K_\eta$ to emphasize its dependence on the conformal exponent
$\eta$, of $(\bfR^2,\e^\eta\delta_{ij})$, is given by
\be\lb{x16}
K_\eta=-\frac12 \e^{-\eta}\Delta\eta.
\ee
Hence (\ref{x15}) simply says
\be\lb{x17}
G_{00}=-G_{33}=-K_\eta;\quad G_{\mu\nu}=0,\quad (\mu,\nu)\neq(0,0)\mbox{ or } (3,3).
\ee
That is, the only nontrivial components $G_{00}$ and $G_{33}$ of the metric element (\ref{x14}) are given by the Gauss curvature of the 2-surface
$(\bfR^2,\e^\eta\delta_{ij})$. In order to achieve consistency in the Einstein equations (\ref{x3}) and maintain the field-theoretical concept that
the temporal component of the energy-momentum tensor $T_{\mu\nu}$, i.e., $T_{00}$, is recognized as the Hamiltonian energy density, usually denoted as
${\cal H}$, we infer from (\ref{x17}) the condition
\be\lb{x18}
T_{00}=-T_{33}={\cal H};\quad T_{\mu\nu}=0,\quad (\mu,\nu)\neq(0,0)\mbox{ or } (3,3).
\ee

{\bf Einstein equations in reduced form versus Nirenberg's problem}

With (\ref{x17}) and (\ref{x18}), we see that the Einstein equations (\ref{x3}) are recast into the single equation,
\be\lb{x19}
K_\eta=8\pi G{\cal H},
\ee
over $(\bfR^2,\e^\eta\delta_{ij})$.
This equation resembles the prescribed curvature equation, $K_g=F$, in Nirenberg's problem, where the Hamiltonian energy density is ``prescribed" by
the matter content of a specific physical model of interest, instead, which makes the statement
\be\lb{x20}
\mbox{Geometry or gravity}=\mbox{Energy or matter content present},
\ee
unambiguously, and will be the focus of our subsequent discussion.

\section{Cosmic strings in exact forms}\lb{sec-x4}
\setcounter{equation}{0}

In this section, we consider two concrete examples of cosmic strings that lay structural foundation of the subject. The first one is due to
Letelier \cite{Let} based on a prescribed Dirac distribution source formalism. The second one is from the construction of Comtet and Gibbons \cite{CG}
based on the $\sig$-model or the harmonic map model.

{\bf Dirac distribution strings}

As the first and simplest example of a multiple string solution to the reduced Einstein equation \eq{x19}, we follow \cite{Let} to take
\be\lb{y1}
{\cal H}=\sum_{s=1}^N\sig_s\delta_{p_s},
\ee
where $\delta_p$ denotes the Dirac distribution concentrated at the point $p\in \bfR^2$ over the 2-surface $(\bfR^2,\e^\eta\delta_{ij})$ and
$\sig_1,\dots,\sig_N>0$ are local string strengths. Thus, through \eq{x19}, the Gauss curvature is given in terms of a sum of the Dirac distributions as well. Inserting \eq{y1} into \eq{x19} and using \eq{x16}, we see that 
\be
\lb{y2}
\eta(x)+4G\sum_{s=1}^N \sig_s\ln|x-p_s|^2,\quad x=(x^1,x^2),
\ee
is a harmonic function over $\bfR^2$, which may be taken to be an arbitrary constant for simplicity. Inserting this fact into \eq{x14}, we obtain
the multiple string metric \cite{Let}
\be\lb{y3}
\dd s^2=\dd t^2 -(\dd x^3)^2-\lm \left(\prod_{s=1}^N |x-p_s|^{2\sig_s}\right)^{-4G}((\dd x^1)^2+(\dd x^2)^2),
\ee
explicitly, where $\lm>0$ is an arbitrary constant. This metric is singular at the string centers, $p_1,\dots,p_N$, where the Gauss curvature
diverges like the Dirac distribution, as prescribed.

Use polar coordinates $(r,\theta)$ to represent the Cartesian coordinates $(x^1,x^2)$. We recall that the flat-spacetime Minkowskian metric reads
\be\lb{y4}
\dd s^2=\dd t^2 -(\dd x^3)^2-((\dd x^1)^2+(\dd x^2)^2)=\dd t^2 -(\dd x^3)^2-\left(\dd r^2+r^2\dd\theta^2\right).
\ee
On the other hand, the multiple string metric \eq{y3} assumes the asymptotic form
\be\lb{y5}
\dd s^2=\dd t^2 -(\dd x^3)^2-r^{-8G\sig}\left(\dd r^2+r^2\dd\theta^2\right),
\ee
for $r=|x|\gg1$,where
$
\sig=\sum_{s=1}^N\sig_s
$
is the total string strength or energy since $\sig=\int_{\bfR^2}{\cal H}\e^\eta\,\dd x$ in view of \eq{y1}. With $G\ll1$, we may assume $4G\sig<1$
and use the change of variables
\be\lb{y6}
\rho=(1-4G\sig)^{-1} r^{1-4G\sig},\quad \omega=(1-4G\sig)\theta,
\ee
 to recast \eq{y5} formally into
\be\lb{y7}
\dd s^2=\dd t^2 -(\dd x^3)^2-\left(\dd \rho^2+\rho^2\dd\omega^2\right),
\ee
which assumes the identical form as \eq{y4} (in polar coordinates). However, such an identification is only valid locally but not globally: As one
travels around the infinity of $\bfR^2$, the polar angle $\theta$ completes a circular round in the full amount $2\pi$ while in correspondence the
``polar angle" $\omega$ registers the value
\be\lb{y8}
\omega=2\pi(1-4G\sig),
\ee
which is less than $2\pi$. The shortage,
\be\lb{y9}
\delta=8\pi G\sig=8\pi G\sum_{s=1}^N\sig_s,
\ee
is called the deficit angle whose presence indicates that, although infinity is preserved but it is no longer flat and is exhibited as a conical singularity. 

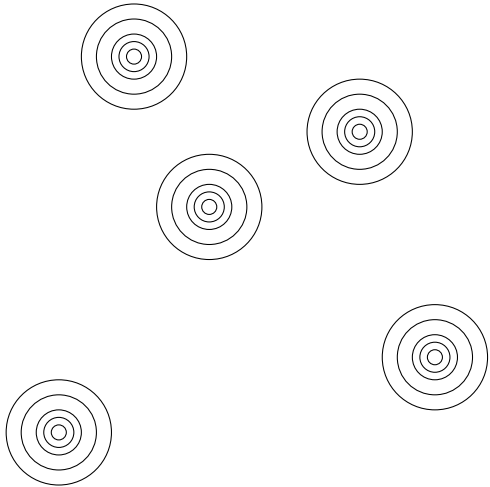
\begin{figure}[H]
\begin{center}
\begin{tikzpicture}
    \draw (0,0) circle [radius=.7cm];
\draw (0,0) circle [radius=.5cm];
\draw (0,0) circle (.3cm);
\draw (0,0) circle (.2cm);
\draw (0,0) circle (.1cm);
 \draw (1,5) circle [radius=.7cm];
\draw (1,5) circle [radius=.5cm];
\draw (1,5) circle (.3cm);
\draw (1,5) circle (.2cm);
\draw (1,5) circle (.1cm);
 \draw (2,3) circle (.7cm);
 \draw (2,3) circle (.5cm);
 \draw (2,3) circle (.3cm);
 \draw (2,3) circle (.2cm);
 \draw (2,3) circle (.1cm);
 \draw (4,4) circle (.7cm);
 \draw (4,4) circle (.5cm);
 \draw (4,4) circle (.3cm);
 \draw (4,4) circle (.2cm);
 \draw (4,4) circle (.1cm);
 \draw (5,1) circle (.5cm);
 \draw (5,1) circle (.7cm);
 \draw (5,1) circle (.3cm);
 \draw (5,1) circle (.2cm);
 \draw (5,1) circle (.1cm);
\end{tikzpicture}
\caption{An illustrative plot of a multicentered distribution of five cosmic strings giving rise to locally concentrated energy density and the associated Gauss
curvature.}\lb{F3}
\end{center}
\end{figure}

In Figure \ref{F3}, we present a plot of the level curves of the energy density or the Gauss curvature of a system of five cosmic strings with identical local
string strengths.

It is interesting to note that, although such cosmic strings behave themselves as local energy and curvature lumps such that
the induced gravitational metric is also singular at those lump centers, these local properties are carried over to infinity as well in
the form of making a conical singularity there as evidenced by the onset of a deficit angle. In particular, the spacetime can never be
asymptotically flat whenever cosmic strings are present. This feature is in sharp contrast against what we know in the presence of a black hole. To
see this, recall for example that the metric element of a Reissner--Nordstr\"{o}m black hole of mass $M$ and charge $Q$ reads
\cite{Car,MTW,Wald}
\be\lb{y10}
\dd s^2=\left(1-\frac{2GM}r+\frac{4\pi G Q^2}{r^2}\right)\dd t^2 - \left(1-\frac{2GM}r+\frac{4\pi G Q^2}{r^2}\right)^{-1}\dd r^2
-r^2\left(\sin^2\varphi\,\dd\theta^2+\dd\varphi^2\right),
\ee
in terms of the spherical coordinates $(r,\theta,\varphi)$.
It is  clear that the black-hole spacetime described by the metric \eq{y10} is asymptotically flat. Consequently, a black hole only influences its 
nearby regions but a cosmic influences infinity as well.

{\bf Light deflection problem}

As in the situation of light deflection around a black hole \cite{MTW,Wald,XY1,XY2}, it will be interesting to formulate and study the light deflection
problem, for a beam of light traveling around a single cosmic string or a collection of several cosmic strings, described by the string metric \eq{y3}.

{\bf Einstein--Hilbert action}

We now embark on a journey of study of cosmic strings generated from field theoretical models. In other words, we shall consider cosmic strings
solutions arising from the Einstein equations coupled with matter-field equations. The starting point of such a formalism is the
Einstein--Hilbert action of the generic form
\be\lb{y11}
S=\int\left(\frac R{16\pi G}+{\cal L}\right)\sqrt{|\det(g_{\mu\nu})|}\,\dd x,
\ee
evaluated over the full domain of spacetime. Varying the gravitational metric tensor $g_{\mu\nu}$ and the matter fields, we arrive at the
coupled governing equations
\be\lb{y12}
G_{\mu\nu}=-8\pi G T_{\mu\nu},\quad \delta{\cal L}=0,
\ee
where the second equation denotes the Euler--Lagrange equation of the matter Lagrangian density $\cal L$ over the gravitational spacetime.

{\bf Harmonic map model}

As a simplest field-theoretical model, we follow Comtet--Gibbons \cite{CG} to consider the $\sig$-model or harmonic map model defined by
\be\lb{y13}
{\cal L}=\frac12\pa_\mu\phi\cdot\pa^\mu\phi=\frac12 g^{\mu\nu} \pa_\mu \phi\cdot \pa_\nu\phi,
\ee
where $\phi=(\phi_1,\phi_2,\phi_3)$ is a map from the spacetime into the unit sphere $S^2$ in $\bfR^3$. The constrained range of $\phi$, $|\phi|^2=\phi_1^2+\phi_2^2+\phi_3^2=1$, renders the problem complicated. In order to overcome this complication, Belavin and Polyakov \cite{BP} use the
stereographic projection, $\phi\mapsto u$, $S^2\mapsto \bfC$, given by
\be\lb{y14}
u=u_1+\ii u_2;\quad u_1=\frac{\phi_1}{1+\phi_3},\quad u_2=\frac{\phi_2}{1+\phi_3},
\ee
from the south pole $(0,0,-1)$ of $S^2$ onto the complex plane which is realized as the equator plane of $S^2$, to simplify the problem. Thus \eq{y13} becomes 
\be\lb{y15}
{\cal L}=\frac2{(1+|u|^2)^2}g^{\mu\nu}\pa_\mu u\pa_\nu\ou.
\ee
In view of \eq{y15}, we see that the equations in \eq{y12} become
\bea
&&G_{\mu\nu}=-8\pi G T_{\mu\nu},\quad T_{\mu\nu}=\frac2{(1+|u|^2)^2}(\pa_\mu u \pa_\nu\ou+\pa_\mu\ou\pa_\nu u)-g_{\mu\nu}{\cal L},\lb{y16}\\
&&\frac1{\sqrt{|\det(g_{\alpha\beta})|}}\pa_\mu\left(\frac{\sqrt{|\det(g_{\alpha\beta})|}\,g^{\mu\nu}}{(1+|u|^2)^2}\pa_\nu u\right)=
-\frac{2u}{(1+|u|^2)^3}\,g^{\mu\nu}\pa_\mu u\pa_\nu\ou.\lb{y17}
\eea

{\bf Useful insight from calculation}

We now return to the cosmic string situation so that the spacetime metric element is given by
\eq{x14} and the complex scalar field $u$ depends only on $x^1, x^2$ as well. Thus, we have
\bea\lb{y18}
{\cal H}&=&T_{00}=-T_{33}=\frac{2\e^{-\eta}}{(1+|u|^2)^2}\left(|\pa_1 u|^2+|\pa_2 u|^2\right)\nn\\
&=&\frac{2\e^{-\eta}}{(1+|u|^2)^2}\left(|\pa_1 u\pm\ii \pa_2 u|^2\pm\ii (\pa_1 u\pa_2\ou-\pa_1\ou\pa_2u)\right).
\eea

In order to recognize the meaning of the decomposition \eq{y18}, we recall that, in this context, $\phi$ may be viewed as a map from $S^2=\bfR^2\cup
\{\infty\}$ into $S^2$ such that its topological degree (the Brouwer degree), written $\deg(\phi)$, is well defined and may be given by
the well-know integral \cite{Gui,Ylabook}
\be\lb{y19}
\deg(\phi)=\frac1{4\pi}\int_{\bfR^2}\phi\cdot(\pa_1\phi\times\pa_2\phi)\,\dd x,
\ee
of a Whitehead \cite{Wh} or Chern--Simons \cite{CS1,CS2} type. On the other hand, from \eq{y14}, we have
\be\lb{y20}
\phi\cdot(\pa_1\phi\times\pa_2\phi)=\frac{2\ii}{(1+|u|^2)^2}(\pa_1 u\pa_2\ou-\pa_1\ou\pa_2u).
\ee
Applying \eq{y19} and \eq{y20} in \eq{y18}, we obtain the following topological energy lower bound
\be\lb{y21}
E=\int_{\bfR^2}{\cal H}\e^\eta\,\dd x
=\int_{\bfR^2}\frac2{(1+|u|^2)^2}|\pa_1 u\pm\ii \pa_2 u|^2 \,\dd x+4\pi|\deg(\phi)|
\geq 4\pi|\deg(\phi)|.
\ee
where we follow the sign convention $|\deg(\phi)|=\pm\deg(\phi)$. It is seen that the lower bound of the energy $E$ expressed by \eq{y21} is
attained if and only if $u$ satisfies the equation
\be\lb{y22}
\pa_1 u\pm\ii \pa_2 u=0,
\ee
which says either $u$ or $\ou$ is a meromorphic function of the complex variable $z=x^1+\ii x^2$. To appreciate this result, we remind ourselves
that the matter equation \eq{y17} assumes the form
\be\lb{y23}
\pa_i\left(\frac{\pa_i u}{(1+|u|^2)^2}\right)=-\frac{2u}{(1+|u|^2)^3}|\nabla u|^2.
\ee
It is direct to examine that \eq{y22} implies \eq{y23}. It will be interesting to establish the opposite implication under certain appropriate
condition, for example, assuming the energy is finite. Thus we have achieved \cite{BP} a significant reduction from the highly nonlinear equation
\eq{y23} into the completely integrable, linear, in fact, the Cauchy--Riemann equations, 
\eq{y22}.

Furthermore, with \eq{y22}, it is straightforward to verify that the energy-momentum tensor $T_{\mu\nu}$ given in \eq{y16} satisfies the property
\eq{x18}. This observation prepares our way to a construction \cite{CG} of multicentered cosmic strings.

Besides, in view of \eq{y22}, we can represent \eq{y18} as a total divergence, namely,
\bea
\e^\eta{\cal H}&=&\pm \frac{2\ii}{(1+|u|^2)^2}(\pa_1 u\pa_2\ou-\pa_1\ou\pa_2u)=\pm J_{12}=\pm(\pa_1 J_2-\pa_2 J_1),\lb{y24}\\
J_i&=&\frac{\ii}{1+|u|^2}(u\pa_i\ou-\ou\pa_i u),\quad i=1,2.\lb{y25}
\eea

This result allows us to compute the energy using Green's theorem directly. To see how, we write the solution to \eq{y22} as a rational
function, observing the plus sign without loss of generality, for the sake of definiteness:
\be\lb{y26}
u(z)=c\prod_{s=1}^N(z-p_s)^{-1}\,\prod_{s=1}^M (z-q_s),\quad c\in\bfC\setminus\{0\}.
\ee
For this function, we see that the topological current density $J_i$ defined in \eq{y25} enjoys the estimate
\be\lb{y27}
J_i=\mbox{O}\left(|z|^{-\delta}\right),\quad\delta=1+2\max\{0,N-M\},\quad |z|\gg1.
\ee
Assume from now on $N>M$.
Therefore, we have
\be\lb{y28}
\int_{\bfR^2} J_{12}\,\dd x=\lim_{r\to\infty}\oint_{|z|=r} J_i\,\dd x^i-\lim_{r\to0}\sum_{s=1}^N \oint_{|z-p_s|=r} J_i\,\dd x^i=4\pi N.
\ee
That is, the number of poles, $N$, of a solution $u$ to \eq{y22}, gives rise to the topological degree of the associated map $\phi:\bfR^2\to S^2$.

The afore-going discussion has been focused on the matter equation \eq{y17}, or specifically \eq{y23}, which is in fact
in its reduced form \eq{y22}, with the explicit solution \eq{y26}. It remains to resolve the Einstein equation \eq{x19}.

{\bf Determination of metric exponent}

We now turn our attention to the Einstein equation \eq{x19} where $\cal H$ following from \eq{y18} assumes form
\be\lb{y29}
{\cal H}=\frac{2\e^{-\eta}}{(1+|u|^2)^2}|\nabla u|^2.
\ee
Thus, by virtue of \eq{x16} and \eq{y29}, we see that \eq{x19} leads to
\be\lb{y30}
-\frac1{16\pi G}\Delta \eta =\frac{2}{(1+|u|^2)^2}|\nabla u|^2.
\ee
It will be instructive to see how to represent the right-hand side of \eq{y30} as a total Laplace. For this purpose, we stay away from the zeros
and poles of $u$ and introduce the quantity $v=\ln|u|^2$ or $|u|^2=\e^v$. With this and \eq{y22}, we have
\be\lb{y31}
|\nabla u|^2=\frac12\,\e^v|\nabla v|^2,
\ee
so that \eq{y30} becomes
\be\lb{y32}
-\frac1{16\pi G}\Delta \eta =\frac{\e^v}{(1+\e^v)^2}|\nabla v|^2.
\ee
Moreover, away from the poles and zeros of $u$, we have 
\be\lb{y33}
\Delta \ln(1+\e^v)=\frac{\e^v\Delta v}{1+\e^v}+\frac{\e^v|\nabla v|^2}{(1+\e^v)^2}.
\ee
Using the fact that
\be\lb{y34}
\Delta v=-4\pi\sum_{s=1}^N\delta_{p_s}(z)+4\pi\sum_{s=1}^M \delta_{q_s}(z),\quad z\in\bfR^2,
\ee
we see that \eq{y33} and \eq{y34} enable us to arrive at the relation
\be\lb{y35}
\frac{\e^v}{(1+\e^v)^2}|\nabla v|^2=\Delta \ln(1+\e^v)+4\pi\sum_{s=1}^N\delta_{p_s}(z),
\ee
over the full $\bfR^2$ in the sense of distributions, in which the zeros of $u$ make no explicit appearance.
As a consequence of \eq{y32} and \eq{y35}, we conclude that the ``quantity"
\be\lb{y36}
\frac{\eta}{16\pi G} +\ln(1+\e^v)+\sum_{s=1}^N\ln|x-p_s|^2,\quad x\in\bfR^2,
\ee
is a harmonic function over $\bfR^2$, which may be taken to be an arbitrary constant again. Therefore, the metric exponent $\eta$ is determined.

{\bf Cosmic strings}

Set \eq{y36} to be a constant. Then we obtain
\be\lb{y37}
\e^\eta=\lm\left(\prod_{s=1}^N |x-p_s|^2\,(1+\e^v)\right)^{-16\pi G},\quad \forall \lm>0.
\ee
Inserting \eq{y26} into \eq{y37}, we obtain the multiple cosmic string metric element \cite{CG}:
\be\lb{y38}
\dd s^2=\dd t^2 -(\dd x^3)^2-\lm \left(\prod_{s=1}^N |x-p_s|^2+|c|^2 \prod_{s=1}^M |x-q_s|^2\right)^{-16\pi G} ((\dd x^1)^2+(\dd x^2)^2).
\ee

There are a few interesting features to be noticed in the expression \eq{y38}.

First, this metric is everywhere regular.

Next, since $N>M$, \eq{y38} asymptotically reads
\be\lb{y39}
\dd s^2=\dd t^2 -(\dd x^3)^2-\lm r^{-32\pi N G}\left(\dd r^2+r^2\dd\theta^2\right),\quad r=|x|\gg1,
\ee
in polar coordinates. Thus, we obtain the deficit angle
\be\lb{y40}
\delta=32\pi^2 N G.
\ee

Then \eq{x19}, \eq{y21}, \eq{y22}, \eq{y28}, and \eq{y40} give us the result
\be
\int_{\bfR^2} K_\eta \e^\eta\,\dd x=32\pi^2 G \,|\deg(\phi)|=\delta.
\ee
That is, in this context, the deficit angle, which is topological as well, is exactly the total Gauss curvature.

Consequently, we see that the expression \eq{x20} can now be expanded to indicate the fact
\be
\mbox{Geometry or gravity}=\mbox{Energy or matter content present}=\mbox{Topology},
\ee
in the context of multiple cosmic strings generated from the harmonic map model.

{\bf Geodesic completeness of cosmic string metric}

Finally, let $p,q$ be two points in $(\bfR^2,\e^\eta\delta_{ij})$. Define the distance
\be\lb{y43}
d(p,q)=\inf\left\{\int_0^T \sqrt{g_{ij}\dot{x}^i\dot{x}^j}\,\dd t=\int_0^T \e^{\frac\eta2}|\dot{x}|\,\dd t\,\bigg|\,x=x(t)\in C^1[0,T], x(0)=p, x(T)=q
\right\},
\ee
often referred to as the geodesic metric over $(\bfR^2,\e^\eta\delta_{ij})$. The surface $(\bfR^2,\e^\eta\delta_{ij})$ is called complete,
or geodesically complete, if \eq{y43} makes $\bfR^2$ a complete metric space. With regard to this notion, the classical Hopf--Rinow--de Rham theorem states
that $(\bfR^2,\e^\eta\delta_{ij})$ is complete if and only if each geodesic over $(\bfR^2,\e^\eta\delta_{ij})$ can be extended to a global geodesic
defined on the entire real line $\bfR$. In such a situation, the infimum in \eq{y43} can always be attained by a minimizing geodesic. This property
is of obvious importance. In the present cosmic string situation, we have the global bounds
\be\lb{y44}
C_1(1+r)^{-32\pi N G}\leq\e^{\eta(x)} \leq C_2(1+r)^{-32\pi N G},\quad x\in\bfR^2,\quad r=|x|,
\ee
which enables us to compare the surface hosting cosmic strings with the radially symmetric one defined by the metric element
\be
\tilde{g}_{ij}=(1+r)^{-32\pi N G}\delta_{ij}.
\ee
 For simplicity, let us consider rays emanating from the origin as geodesics. Then the completeness of $\tilde{g}_{ij}$ is recast into the question
whether the geodesic distance from the origin to infinity of $\bfR^2$ is infinite. This is equivalent to asking whether the integral
\be
\int_0^\infty (1+r)^{-16\pi N G}\,\dd r
\ee
is divergent. As a consequence, we arrive at the condition
\be\lb{y46}
N\leq\frac1{16\pi G},
\ee
as a necessary and sufficient condition for the geodesic completeness of the cosmic string metric.

\section{Cosmic strings in Abelian Higgs models}
\setcounter{equation}{0}

In the previous section, we presented multiple cosmic string solutions
of the Einstein equations generated from prescribed Dirac distribution sources and a coupled harmonic map model. These solutions are exact,
explicit, and provide precise and detailed description and understanding of the gravitational system. In this section, we consider more realistic
models involving matter-mediating fields, namely, electromagnetic fields, in the formalism of the classical Abelian Higgs models. In this context,
we again use $u$ to denote a complex scalar field. Let $A_\mu$ denote a real-valued gauge vector field so that $F_{\mu\nu}=\pa_\mu A_\nu
-\pa_\nu A_\mu$ represents the induced electromagnetic field. The associated gauge-covariant derivative reads
\be\lb{z1}
D_\mu u=\pa_\mu u-\ii A_\mu u,\quad \mu=0,1,2,3,
\ee
giving rise to the commutator relation
\be\lb{z2}
[D_\mu,D_\nu] u=(D_\mu D_\nu-D_\nu D_\mu) u=-\ii F_{\mu\nu}u,
\ee
so that $F_{\mu\nu}$ arises as well as a ``curvature" quantity as in the Riemannian situation. 

{\bf Classical Abelian Higgs theory}

In this case, the Lagrangian action density assumes the form \cite{JT}
\be\lb{z3}
{\cal L}=-\frac14 g^{\mu\mu'}g^{\nu\nu'}F_{\mu\nu}F_{\mu'\nu'}+\frac12 g^{\mu\nu}D_\mu u \overline{D_\nu u}-\frac18(|u|^2-1)^2.
\ee
Multiple cosmic strings of this model have been constructed \cite{Ycs1,Ycs2} based on the formalism in \cite{CG}.

{\bf Abelian Higgs model hosting strings and antistrings}

Use ${\bf n}$ to denote the north pole of $S^2$, ${\bf n}=(0,0,1)$, and $\phi$ a map from the spacetime into $S^2$. Based on the gauged harmonic map
model of Schroers \cite{Sch1,Sch2}, we consider the Lagrangian action density
\be\lb{z4}
{\cal L}=-\frac14 g^{\mu\mu'}g^{\nu\nu'}F_{\mu\nu}F_{\mu'\nu'}+\frac12 g^{\mu\nu}(D_\mu \phi)\cdot ({D_\nu \phi})-\frac12({\bf n}\cdot\phi)^2,
\ee
where $D_\mu\phi=\pa_\mu\phi-A_\mu({\bf n}\times\phi)$. With the complexification \eq{y14}, we come up with the modified Lagrangian action density
\be\lb{z5}
{\cal L}=-\frac14 g^{\mu\mu'}g^{\nu\nu'}F_{\mu\nu}F_{\mu'\nu'}+\frac2{(1+|u|^2)^2} g^{\mu\nu}D_\mu u \overline{D_\nu u}-\frac12\left(\frac{|u|^2-1}{|u|^2+1}\right)^2.
\ee
This theory has a few distinctive features worthy noticing. The first one is that both zeros and poles are allowed in the theory which make
equal and indistinguishable contributions to the matter energy and gravitational fine structures of the problem \cite{Yvav1,Yvav2,Yvav3}.
The second one is that, in addition to its usual gauge invariance, it also enjoys a flipping symmetry, given by
\be
(u,A_\mu)\to \left(\frac1u,-A_\mu\right),
\ee
which explains why the zeros and poles of $u$ play equal roles in the model. Moreover, it is seen that \eq{z5} returns to \eq{z3} in the limit
$|u|^2\approx 1$ and that in both \eq{z3} and \eq{z5} there is a spontaneously broken $U(1)$ gauge symmetry which is responsible for the
presence of vortices and strings of similar magnetic and topological origins \cite{SSY}.

{\bf Generalized Abelian Higgs theory}

In the rest of this section, we present an Abelian Higgs theory aimed at coupling with Einstein's general relativity which may be used to
generate some new families of multiple cosmic string solutions along the formulation \cite{CG} and developments \cite{Ycs1,Ycs2,Yvav1,Yvav2,Yvav3}.

This part of the study is of two main purposes: (i) To unify the models \eq{z3} and \eq{z5} in a general setting. (ii) To introduce some new families
of nonlinear partial differential equations of theoretical physics interests in the spirit of Nirenberg's problem.

To proceed, we follow Lohe's study \cite{Lo1} (see also \cite{Lo2,Lo3,WY} for subsequent mathematical work) to consider
the Lagrangian action density
\be\lb{z7}
{\cal L}=-\frac14 g^{\mu\mu'}g^{\nu\nu'}F_{\mu\nu}F_{\mu'\nu'}+\frac12 F(s) \,g^{\mu\nu}D_\mu u \overline{D_\nu u}-\frac12 w^2(s),
\ee
where $F$ and $w$ are some real-valued functions in the variable $s=|u|^2$ to be specified to achieve the desired Bogomol'nyi structure \cite{Bo}
in the model. From
\eq{z7}, we obtain the associated energy-momentum tensor
\be\lb{z8}
T_{\mu\nu}=-g^{\mu'\nu'}F_{\mu\mu'}F_{\nu\nu'}+\frac12 F(s)(D_\mu u\overline{D_\nu u}+\overline{D_\mu u} D_\nu u)-g_{\mu\nu}{\cal L}.
\ee
With \eq{z8}, it can be calculated from the Einstein--Hilbert action \eq{y11} that the full coupled Einstein and matter field equations are
\bea
&&G_{\mu\nu}=-8\pi G T_{\mu\nu},\\
&& \frac1{\sqrt{|\det(g_{\alpha\beta})|}}\pa_\mu\left(F(s){\sqrt{|\det(g_{\alpha\beta})|}\,g^{\mu\nu}}\pa_\nu u\right)=(F'(s) g^{\mu\nu}D_\mu u\overline{D_\nu u}-2w(s)w'(s))u,\\
&& \frac1{\sqrt{|\det(g_{\alpha\beta})|}}\pa_{\mu'}\left(g^{\mu\nu}g^{\mu'\nu'}\sqrt{|\det(g_{\alpha\beta})|} F_{\nu\nu'}\right)
={F(s)}\frac
\ii2 g^{\mu\nu}(u\overline{D_\nu u}-\ou D_\nu u).
\eea

On the other hand, with the string metric \eq{x14} and staying within the static and axial symmetric situation as before, we obtain from \eq{z8}
the Hamiltonian energy density
\bea\label{z12}
{\cal H}&=&T_{00}=-T_{33}=-{\cal L}\nn\\
&=&\frac12\e^{-2\eta}F_{12}^2+\frac12 F(s) \e^{-\eta}(|D_1 u|^2+|D_2 u|^2)+\frac12 w^2(s)\nn\\
&=&\frac12(\e^{-\eta}F_{12}\mp w)^2\pm \e^{-\eta} F_{12} w+\frac{F(s)}2\e^{-\eta}(|D_1 u\pm\ii D_2 u|^2\pm\ii [D_1 u \overline{D_2 u}-\overline{D_1 u} D_2 u])\nn\\
&=&\frac12([\e^{-\eta}F_{12}\mp w]^2+F(s) \e^{-\eta}|D_1 u\pm\ii D_2 u|^2)\pm\frac\tau2 \e^{-\eta} F_{12}\nn\\
&& \pm \frac12\e^{-\eta}([2w-\tau]F_{12}+F(s)\ii [D_1 u \overline{D_2 u}-\overline{D_1 u} D_2 u]),
\eea
where we have regrouped the terms on the right-hand side of \eq{z12} such that the first term consists of quadratures, the second term is topological, giving rise to the first Chern class density, where the parameter $\tau$ is inserted for manipulation, and the last term is to be recognized as the total divergence of a current density,
which is what we now do.

To this end, and suggested by \eq{y24} and \eq{y25}, we form the current density
\be\lb{z13}
J_i=f(s)\frac\ii2(u\overline{D_i u}-\ou D_i u),\quad i=1,2.
\ee
With \eq{z13} and using \eq{z2}, we have
\be\lb{z14}
J_{12}=\pa_1 J_2-\pa_2 J_1=-f(s)s F_{12}+(f(s)+f'(s) s) \ii (D_1 u \overline{D_2 u}-\overline{D_1 u} D_2 u).
\ee
Identify \eq{z14} with the last term on the right-hand side of \eq{z12}, we get the relations
\be\lb{z15}
w(s)-\frac\tau2=-f(s)s,\quad F(s)=2(f(s)+f'(s)s).
\ee
Therefore we arrive at the conclusion
\be\lb{z16}
w(s)=\frac12\int_s^1 F(\rho)\,\dd\rho\quad\mbox{or equivalently}\quad F(s)=-2w'(s), \quad w(1)=0.
\ee

As an illustration, we consider the examples
\bea
&&\tau=1, \quad f(s)=\frac12, \quad w(s)=\frac{1-s}2,\quad F(s)=1;\lb{z17}\\
&&\tau=2, \quad f(s)=\frac
2{1+s},\quad w(s)=\frac{1-s}{1+s},\quad F(s)=\frac4{(1+s)^2}.\lb{z18}
\eea
It is clear that \eq{z17} leads to the model \eq{z3} and \eq{z18} recovers \eq{z5}. 
In general, the function $f(s)$ in \eq{z13} plays the role of
an intermediate auxiliary variable and the key relation is given by \eq{z16}. 

For simplicity, we assume that $u$ has no poles so that the matter and gauge fields, $u$ and $A_1, A_2$ involved, are regular. From \eq{z16}, we see
that we have the normalization $w(1)=0$ such that the boundary condition
\be
|u|^2=1
\ee
is to be imposed at infinity. This condition and the first equation in \eq{z15} lead to the condition
\be
f(1)=\frac\tau2.
\ee
Thus $f(|u|^2)$ is nonvanishing at infinity. From \eq{z13}, we see that we may require that $D_i u$ vanish at infinity sufficiently rapidly, which is
consistent with the form of the Hamiltonian $\cal H$ stated in \eq{z12}, which ensures that $J_i$ vanishes at infinity sufficiently rapidly. For our study, we shall assume
\be
J_i =\mbox{O}(|x|^{-\delta}),\quad \delta>1,\quad |x|\gg1,
\ee
which renders us the vanishing result
\be\lb{z22}
\int_{\bfR^2}J_{12}\,\dd x=\lim_{r\to\infty}\oint_{|x|=r} J_i\,\dd x^i=0.
\ee
Integrating \eq{z12} over $(\bfR^2,\e^{\eta}\delta_{ij})$ and inserting \eq{z22}, we have
\be\lb{z23}
E=\int_{\bfR^2}{\cal H}\e^\eta\,\dd x\geq \tau\pi N,\quad \int_{\bfR^2}F_{12}\,\dd x=\pm 2\pi N,
\ee
where $\pm N$ is usually a topological integer defined as the first Chern class of the model such that $N$ represents the number of zeros 
of $u$ or the total string number. The energy lower bound is saturated if and only if the two quadratic terms in \eq{z12} identically vanish:
\bea
F_{12}&=&\pm \e^\eta w(|u|^2),\lb{z24}\\
D_1 u\pm\ii D_2 u&=&0.\lb{z25}
\eea
The equation \eq{z25} is the gauge-covariant extension of the Cauchy--Riemann equation \eq{y22} which may be rewritten as 
\be\lb{z26}
(\pa_1\pm\ii\pa_2)u=\ii(A_1\pm\ii A_2)u.
\ee
Resolving \eq{z26}, we see that, away from the zeros of $u$, there holds the identity
\be\lb{z27}
F_{12}=\mp\frac12\Delta \ln|u|^2.
\ee
On the other hand, \eq{z25} or \eq{z26} and the $\overline{\pa}$-Poincar\'{e} lemma \cite{JT} indicates that the zeros of $u$ are all discrete and
of integer multiplicities. Let the zeros of $u$ be $p_1,\dots,p_N$ (counting multiplicities). Then we can combine \eq{z24} and \eq{z27} to arrive
at the governing equation
\be\lb{z28}
\Delta v=-2\e^\eta w(\e^v)+4\pi\sum_{s=1}^N \delta_{p_s}(x),\quad x\in\bfR^2,
\ee
where $|u|^2=\e^v$ and the Dirac function $\delta_p(x)$ is defined over the usual plane $\bfR^2$. Moreover, with \eq{z24} and \eq{z25}, we can 
verify as before that $T_{\mu\nu}$ defined by \eq{z8} satisfies $T_{\mu\nu}=0$ for $(\mu,\nu)\neq(0,0)$ or $(3,3)$. Hence we arrive at the
Gauss curvature equation \eq{x19} again.

{\bf Multiple vortex equations}

Of independent interest is the situation when gravity is absent, characterized by setting $G=0$ in \eq{x19}. Hence we may take $\eta$ to be trivial,
$\eta=0$, and the equation \eq{z28} becomes the following Liouville type equation:
\be\lb{z29}
\Delta v=-2 w(\e^v)+4\pi\sum_{s=1}^N \delta_{p_s}(x),\quad x\in\bfR^2.
\ee
Besides those potential profiles for $w(s)$ given in \eq{z17} and \eq{z18}, some other forms of $w(s)$ of interest are
\bea
&&w(s)=\frac{1-s^m}{2m},\quad F(s)=s^{m-1},\quad m=1,2,\dots,\lb{z30}\\
&&w(s)=\frac12\ln\left(\frac{1+b}{s +b}\right),\quad F(s)=\frac{1}{s+b},\quad b>0,\lb{z31}\\
&&w(s)=\frac\alpha2\left(\frac\pi4-\arctan s\right),\quad F(s)=\frac\alpha{1+s^2},\quad\alpha>0,\lb{z32}\\
&&w(s)=\frac{1-s^m}{1+s^m},\quad F(s)=\frac{4m s^{m-1}}{(1+s^m)^2},\quad m=1,2,\dots,\lb{zx33}\\
&&w(s)=\frac\beta2\left(1-\frac{\sinh s}{\sinh 1}\right),\quad F(s)=\frac{\beta\cosh s}{\sinh 1},\quad \beta>0.\lb{zx34}
\eea
The cases \eq{z30} and \eq{z31} over the full $\bfR^2$ have been considered in \cite{WY}. The case \eq{z32} is a new model. The case \eq{zx33}
when $m=1$ is the model \eq{z5}. The case \eq{zx33} when $m\geq2$ and the case \eq{zx34} are new.

In view of \cite{Aub,Bra,CY,Nog,SSY,WY2}, it will also be interesting to study
\eq{z28} over a compact surface, $(S,g)$, in particular, a flat torus. For this purpose, let $|S|$ denote the surface area of $S$ with respect to the metric element $g$.
Over $(S,g)$, the equation \eq{z29} becomes
\be\lb{z33}
\Delta_g v=-2 w(\e^v)+4\pi\sum_{s=1}^N \delta_{p_s}(x),\quad x\in S.
\ee
Use $v_0$ to denote a source function over $S$ satisfying \cite{Aub}
\be\lb{z34}
\Delta_g v_0=-\frac{4\pi N}{|S|}+4\pi\sum_{s=1}^N \delta_{p_s}(x),\quad x\in S.
\ee
Represent the solution $v$ to \eq{z33} as $v=v_0+V$. Then $V$ satisfies the equation
\be\lb{z35}
\Delta_g V=-2w(\e^{v_0+V})+\frac{4\pi N}{|S|}.
\ee
This equation resembles the equation \eq{x5} in Nirenberg's problem. Integrating \eq{z35}, we get the topological constraint
\be
\int_S w(\e^{v_0+V})\,\dd\Omega_g=2\pi N,
\ee
resembling \eq{x8}. Hence the solvability of \eq{z35} is equivalent to that of the constrained minimization problem
\be\lb{z37}
\min\left\{\int_S\left(\frac12|\nabla V|^2+\frac{4\pi N}{|S|} V\right)\dd\Omega_{g}\,\bigg|\int_S w(\e^{v_0+V})\, \dd\Omega_{g}=2\pi N\right\}.
\ee
Of course, other elliptic methods such as the fixed-point theory,  sub- and super-solution iterations, and heat-flow approach may be effective here too.

{\bf Cosmic string equations}

As in the situation of harmonic map model, in the present situation, it is crucial to realize the quantity $\e^\eta{\cal H}$ as a total 
Laplace such that we may resolve the Einstein equation \eq{x19} to obtain the conformal factor $\e^\eta$ in terms of the unknown $v$ in \eq{z28}.
To this end, we first use \eq{z24} to get
\be\lb{z38}
\e^{-\eta}F_{12}^2 +\e^\eta w^2=\pm 2 F_{12} w.
\ee
Besides, as a consequence of \eq{z25}, we may extend the identity \eq{y31} to get
\be\lb{z39}
|D_1 u|^2+|D_2 u|^2=\frac12 \e^v|\nabla v|^2.
\ee
From \eq{z38}, \eq{z39}, and using \eq{z15} and \eq{z27}, we obtain
\bea\lb{z40}
2\e^\eta{\cal H}&=&\e^{-\eta}F_{12}^2+ F(s)(|D_1 u|^2+|D_2 u|^2)+ \e^\eta w^2(s)\nn\\
&=&-\left(\frac\tau2 -f(s) s\right)\Delta v+(f(s)+f'(s) s)\e^v|\nabla v|^2\nn\\
&=& f(\e^v)\e^v\Delta v+(f(\e^v)+f'(\e^v)\e^v)\e^v|\nabla v|^2 -\frac\tau2\Delta v\nn\\
&=&\Delta h(\e^u)-\frac\tau2\Delta v,\quad h(s)=\int_0^s f(\rho)\,\dd\rho,
\eea
away from the zeros of $u$. When considering the full plane $\bfR^2$, the source term $4\pi \sum_{s=1}^N \delta_{p_s}(x)$ resulting from
the presence of the zeros $p_1,\dots,p_N$ of $u$ (without the presence of poles), arising from $\Delta v$,
should be added to the right-hand side of \eq{z40} so that the left-hand side of \eq{z40} stays regular. With this observation and returning
to \eq{x19} with \eq{x16}, we conclude that the quantity
\be
\frac\eta{8\pi G}+h(\e^v)-\frac{\tau v}2 +\frac\tau2\sum_{s=1}^N\ln|x-p_s|^2
\ee
is harmonic which may be taken to be an arbitrary constant. Consequently, we obtain the gravitational metric factor
\be\lb{z42}
\e^\eta =\lm\left(\frac{\e^{\tau v}}{\e^{2h(\e^v)}}\prod_{s=1}^N|x-p_s|^{-2\tau}\right)^{4\pi G}=\mbox{O}(|x|^{-8\pi\tau N G}),\quad |x|\gg1,
\ee
since $v$ vanishes at infinity. This result leads to the deficit angle 
\be\lb{845}
\delta=8\pi^2\tau N G.
\ee
 On the other hand, recall that \eq{z23}  gives us the total energy
\be
E=\pi\tau N.
\ee
 Thus, integrating \eq{x19} renders the total curvature 
\be\lb{846}
\int_{\bfR^2}K_\eta\e^\eta\dd x=8\pi^2\tau NG,
\ee
 in agreement with the deficit angle
again. Moreover, in line with our study in Section \ref{sec-x4}, we arrive at the condition 
\be\lb{847}
N\leq \frac1{4\pi\tau G}
\ee
 for the geodesic completeness of
the cosmic string metric. Note that the results \eq{845}--\eq{847} clearly illustrate the roles played by the parameter $\tau$, geometrically,
energetically, and topologically.

Substituting \eq{z42} into \eq{z28}, with \eq{z15}, we finally obtain the cosmic string equation
\be\lb{z43}
\Delta v= \lm\left({\e^{\tau v}}\prod_{s=1}^N|x-p_s|^{-2\tau}\exp\left[-2\int_0^{\e^v} f(\rho)\,\dd\rho\right]\right)^{4\pi G}
 \left(2f(\e^v)\e^v-\tau\right)+4\pi\sum_{s=1}^N \delta_{p_s}(x),\quad x\in\bfR^2.
\ee
In the classical situation \cite{CG,Ycs1,Ycs2}, we have $\tau=1$ and that the function $f(s)$ is $f(s)=\frac12$ (cf. \eq{z17}). Then \eq{z43} becomes
\be\lb{z44}
\Delta v= \lm\left({\e^{v-\e^v}}\prod_{s=1}^N|x-p_s|^{-2}\right)^{4\pi G}
 \left(\e^v-1\right)+4\pi\sum_{s=1}^N \delta_{p_s}(x),\quad x\in\bfR^2,
\ee
after an update of the free coupling parameter $\lm$. Although the equations \eq{z43} and \eq{z44} appear complicated, the study in \cite{Ycs2}
indicates that they could be understood by considering the solutions of the equations without gravity, namely \eq{z29} or \eq{z43} and \eq{z44} with
setting $G=0$, and manipulating the free parameter $\lm$. See also \cite{Yvav2} for some similar methods for different problems.

\section{Other extended models}
\setcounter{equation}{0}

It will also be interesting to study multiple cosmic strings arising from other extended Abelian Higgs theories. These include the model
defined by the Lagrangian action density \cite{Con,Izq}
\be\lb{z45}
{\cal L}=-\frac14 F(s) g^{\mu\mu'}g^{\nu\nu'}F_{\mu\nu}F_{\mu'\nu'}+\frac12 \,g^{\mu\nu}D_\mu u \overline{D_\nu u}-V(s),
\ee
 that of the Born--Infeld theory type \cite{Cas,Han}
\be\lb{z46}
{\cal L}=\frac1\beta\left(1-\sqrt{1+\frac\beta2 F(s) g^{\mu\mu'}g^{\nu\nu'}F_{\mu\nu}F_{\mu'\nu'}}\right)+\frac12 G(s)\,g^{\mu\nu}D_\mu u \overline{D_\nu u}-V(s),
\ee
and subsequently that of the combined Lagrangian action density
\be\lb{z45x}
{\cal L}=-\frac14 F(s) g^{\mu\mu'}g^{\nu\nu'}F_{\mu\nu}F_{\mu'\nu'}+\frac12 G(s) g^{\mu\nu}D_\mu u \overline{D_\nu u}-V(s),
\ee
where $F(s), G(s), V(s)$ are some suitable functions of $s=|u|^2$ to be determined to achieve a Bogomol'nyi structure \cite{Bo} and the condition \eq{x18}. Below we elaborate on \eq{z45} briefly.

{\bf Hamiltonian calculation and decomposition}

With $F(s)=h^2(s)$ and $V(s)=\frac12 w^2(s)$, \eq{z45} becomes
\be\lb{z53}
{\cal L}=-\frac14 h^2(s) g^{\mu\mu'}g^{\nu\nu'}F_{\mu\nu}F_{\mu'\nu'}+\frac12 \,g^{\mu\nu}D_\mu u \overline{D_\nu u}-\frac12 w^2(s),
\ee
whose energy-momentum tensor reads
\be\lb{z54}
T_{\mu\nu}=-h^2(s)g^{\mu'\nu'}F_{\mu\mu'}F_{\nu\nu'}+\frac12(D_\mu u\overline{D_\nu u}+\overline{D_\mu u} D_\nu u)-g_{\mu\nu}{\cal L}.
\ee
Observing the same static string assumption such that the fields depend on $x^1,x^2$ only and $A_0=A_3=0$ and that the string metric
is given by \eq{x14}, we obtain from \eq{z54} the Hamiltonian energy density
\bea\lb{z55}
{\cal H}&=&T_{00}=-T_{33}=-{\cal L}\nn\\
&=&\frac12 \e^{-2\eta} h^2(s) F_{12}^2+\frac12 \e^{-\eta}(|D_1 u|^2+|D_2 u|^2)+\frac12 w^2(s)\nn\\
&=&\frac12\left((\e^{-\eta} h(s) F_{12}\mp w(s))^2+\e^{-\eta}|D_1u\pm\ii D_2 u|^2\right)\pm\frac12 \e^{-\eta}F_{12}\nn\\
&&\pm\frac12 \e^{-\eta}\left((2h(s)w(s)-1)F_{12}+\ii (D_1u \overline{D_2 u}-\overline{D_1u}D_2u)\right).
\eea
On the right-hand side of \eq{z55}, the terms are grouped into three quantities: The first one consists of quadratic terms, the second is
topological, and the third may be recast into a total divergence as before under some conditions on $h(s)$ and $w(s)$. To see how, we note 
from \eq{z13} and \eq{z14} that, if we set
\be\lb{z56}
2h(s) w(s)=1-s,
\ee
then the right-hand side of \eq{z55} becomes
\be
{\cal H}=\frac12\left((\e^{-\eta} h(s) F_{12}\mp w(s))^2+\e^{-\eta}|D_1u\pm\ii D_2 u|^2\right)\pm\frac12 \e^{-\eta}F_{12}\pm\frac12\e^{-\eta}J_{12},
\ee\lb{z57}
where $J_i$ is given by \eq{z13} with $f(s)=1$. 

{\bf Bogomol'nyi equations}

Hence we derive the Bogomol'nyi topological lower bound $E=\int {\cal H}\e^{\eta}\,\dd x\geq\pi N$ where
$N$ is the total string number and this lower bound is saturated by the equations
\bea
h(|u|^2) F_{12}&=&\pm \e^\eta w(|u|^2),\lb{z58}\\
D_1u\pm\ii D_2u&=&0.\lb{z59}
\eea
As a consequence, we see that $T_{\mu\nu}$ given in \eq{z53} satisfies $T_{\mu\nu}=0$ when $(\mu,\nu)\neq (0,0)$ or $(3,3)$, as before.
Thus, using \eq{z27} and setting $v=\ln|u|^2$, we obtain from \eq{z58} and \eq{z59}, while observing \eq{z56}, the equation
\be\lb{z60}
\Delta v={\e^\eta}\frac{(\e^v-1)}{h^2(\e^v)}+4\pi\sum_{s=1}^N \delta_{p_s}(x),\quad x\in\bfR^2,
\ee
which is similar to \eq{z28}.

{\bf Metric factor}

To resolve the Einstein equation \eq{x19}, we use \eq{z58} to represent $\cal H$ given in the second line of \eq{z55} as
\be\lb{z61}
{\cal H}=\pm\e^{-\eta} h(s) F_{12} w+\frac12\e^{-\eta}(|D_1 u|^2+|D_2 u|^2).
\ee
In view of \eq{z56}, \eq{z39}, and \eq{z27}, we may recast \eq{z61} to get
\be\lb{z62}
4\e^\eta{\cal H}=(\e^v-1)\Delta v+\e^v|\nabla v|^2,
\ee
away from the zeros $p_1,\dots,p_N$ of $u$. Taking account of these zeros, \eq{z62} leads to
\be\lb{z63}
4\e^\eta{\cal H}=\Delta (\e^v-v)+4\pi\sum_{s=1}^N\delta_{p_s}(x),\quad x\in\bfR^2.
\ee
Therefore, in view of \eq{x16}, \eq{x19}, and \eq{z63}, we conclude again that the quantity
\be\lb{z64}
\frac\eta{4\pi G}+\e^v-v+\sum_{s=1}^N\ln|x-p_s|^2
\ee
is a harmonic function, which may be taken to be an arbitrary constant. Thus, we find the gravitational metric factor:
\be\lb{z65}
\e^\eta=\lm\left(\e^{v-\e^v}\prod_{s=1}^N |x-p_s|^{-2}\right)^{4\pi G},\quad \lm>0.
\ee

{\bf Multiple string equation}

Inserting \eq{z65} into \eq{z60}, we obtain the governing equation for $N$ prescribed cosmic strings located at $p_1,\dots,p_N$:
\be\lb{z66}
\Delta v= \lm\left(\e^{v-\e^v}\prod_{s=1}^N |x-p_s|^{-2}\right)^{4\pi G}\frac{(\e^v-1)}{h^2(\e^v)}+4\pi\sum_{s=1}^N \delta_{p_s}(x),\quad x\in\bfR^2,
\ee
where $\lm>0$ is a constant which may be taken to be arbitrarily large.

{\bf Two examples}

We now examine a few concrete examples. 

Taking $h(s)=1$, the condition \eq{z56} gives us $w(s)=\frac12(1-s)$ so that we return to the classical Abelian Higgs model for which the cosmic
string equation is given by \eq{z44}, which is seen to be contained in \eq{z66} as a special case, indeed, when $h(s)=1$.

Let $\kappa>0$ be a constant and take 
\be\lb{z67}
w(s)=\frac{1}\kappa \sqrt{s}(1-s).
\ee
Then \eq{z56} gives us
\be\lb{z68}
h(s)=\frac{\kappa}{2\sqrt{s}}.
\ee
Inserting \eq{z67} and \eq{z68} into the second line in \eq{z55}, we obtain
\be\lb{z69}
{\cal H}=\frac{\kappa^2}8 \e^{-2\eta} \frac{F_{12}^2}{|u|^2}+\frac12 \e^{-\eta}(|D_1 u|^2+|D_2 u|^2)+\frac1{2\kappa^2} |u|^2(1-|u|^2)^2.
\ee
Without gravity, or $G=0$ or $\eta=0$, \eq{z69} is the Hamiltonian energy density of the self-dual Chern--Simons--Higgs theory \cite{HKP,JW}
describing multiply distributed electrically and magnetically charged vortices following the equation
\be\lb{z70}
\Delta v= \lm\e^v(\e^v-1)+4\pi\sum_{s=1}^N \delta_{p_s}(x),\quad x\in\bfR^2,
\ee
with suppressed parameters, which has been well studied in \cite{CY,Chae,SY1,SY2,Tar}. With gravity, we obtain from \eq{z66} and \eq{z68} the
equation
\be\lb{z71}
\Delta v= \lm\left(\e^{v-\e^v}\prod_{s=1}^N |x-p_s|^{-2}\right)^{4\pi G}\e^v(\e^v-1)+4\pi\sum_{s=1}^N \delta_{p_s}(x),\quad x\in\bfR^2,
\ee
which is a new problem.

{\bf Some new multiple vortex equations}

To see the structure of this new family of problems more transparently, we consider the situation when gravity is again absent in 
\eq{z66} with $G=0$. From \eq{z56}, we may come up with the examples
\bea
&&h(s)=\frac\kappa{2s^{\frac m2}},\quad w(s)=\frac{s^{\frac m2}(1-s)}\kappa,\lb{z72}\\ 
&&h(s)=\frac\kappa{2(1+s)^{\frac m2}},\quad w(s)=\frac{(1+s)^{\frac m2}(1-s)}\kappa,\lb{z73}\\
&&h(s)=\frac\kappa2(1+s)^{\frac m2},\quad w(s)=\frac{1-s}{\kappa(1+s)^{\frac m2}},\lb{z74}
\eea
where $m\geq1$ is a parameter. These examples give us the respective multiple vortex equations:
\bea
\Delta v&=& \lm \e^{mv} (\e^v-1)+4\pi\sum_{s=1}^N \delta_{p_s}(x),\lb{z75}\\
\Delta v&=& \lm (1+\e^v)^m (\e^v-1)+4\pi\sum_{s=1}^N \delta_{p_s}(x),\lb{z76}\\
\Delta v&=&\lm\frac{(\e^v-1)}{(1+\e^v)^m}+4\pi\sum_{s=1}^N \delta_{p_s}(x),\lb{z77}
\eea
over $\bfR^2$ or a compact surface, where $\lm>0$ is a parameter. The equations \eq{z75} and \eq{z76} extend the Chern--Simons vortex equation
\eq{z70} in different ways and \eq{z77} generalizes the multiple vortex equation in gauged harmonic model based on \eq{z5}. These equations enjoy
some nice analytic structures. For example, $v^+=0$ is a supersolution to \eq{z77}. Now, to approach \eq{z77}, we consider a simplified equation,
say,
\be
\Delta v=\frac\lm{2^m}{(\e^v-1)}+4\pi\sum_{s=1}^N \delta_{p_s}(x),\lb{1029}
\ee
whose solution vanishing at infinity stays negative, $v<0$. For this function $v$, set $v^-=v$. Then we have
\be
\Delta v^->\lm\frac{(\e^{v^-}-1)}{(1+\e^{v^-})^m}+4\pi\sum_{s=1}^N \delta_{p_s}(x).\lb{1030}
\ee
That is, $v^-$ is a subsolution to \eq{z77}. Since $v^+>v^-$, we obtain a solution $v$ to \eq{z77} which vanishes at infinity and satisfies $v^-<v<v^+=0$ pointwise.

See also Manton \cite{Man} and Gudnason \cite{Gud} for some recent reviews on vortex equations.

{\bf Cosmic string equations of interest}

Naturally, the cosmic string
equations of these models, that is, the equation \eq{z66} with \eq{z72}--\eq{z74}, respectively, for example, will be of interest to study as well.

\medskip

In summary, we have seen that cosmic string studies reduce the Einstein tensor into a much reduced form of the tensor given by the Gauss
curvature of a gravitational surface that hosts such strings. At the centers of the strings, both the Gauss curvature and matter energy density
assume locally peaked values and cause a conical singularity at infinity. In all situations studied, the total Gauss curvature is exactly the
deficit angle of the conical singularity which is also determined by the total number of strings as a topological quantity. 
Mathematically, the studies of such cosmic strings depend on understanding some highly nonlinear partial differential equations of the type
of Nirenberg's problem in geometric analysis.

\medskip

The author also dedicates this article to the memory of Professors Chen Shunqing, Guo Penglei, Liu Guangyao, Liu Yaxing, and Zhao Hongxun, whose
encouragement and guidance, given to the author while he was a mathematics undergraduate student during 1978--1982, were enlightening and
impactful.

\medskip

This article was written based on two sequences of lectures
given at School of Mathematics and Statistics, Henan University. The author thanks Professor Xiaosen Han for arranging these
lectures and for suggesting to write this article to commemorate the centennial occasion.

\end{document}